# Report on the XBase Project

E. Zirintsis, G.N.C. Kirby, A. Dearle and R. Morrison
*{vangelis, graham, al, ron }@dcs.st-and.ac.uk*

School of Computer Science, University of St Andrews,
North Haugh, St Andrews KY16 9SS, Scotland

## Abstract

This project addressed the conceptual fundamentals of data storage, investigating techniques for provision of highly generic storage facilities that can be tailored to produce various individually customised storage infrastructures, compliant to the needs of particular applications. This requires the separation of mechanism and policy wherever possible. Aspirations include: actors, whether users or individual processes, should be able to bind to, update and manipulate data and programs transparently with respect to their respective locations; programs should be expressed independently of the storage and network technology involved in their execution; storage facilities should be structure-neutral so that actors can impose multiple interpretations over information, simultaneously and safely; information should not be discarded so that arbitrary historical views are supported; raw stored information should be open to all; where security restrictions on its use are required this should be achieved using cryptographic techniques. The key advances of the research were:

1) the identification of a candidate set of minimal storage system building blocks, which are sufficiently simple to avoid encapsulating policy where it cannot be customised by applications, and composable to build highly flexible storage architectures

2) insight into the nature of append-only storage components, and the issues arising from their application to common storage use-cases

## 1 Introduction

The XBase project aims to provide actors (processes, users) general storage facilities which will be compliant to the needs of particular applications. The main goal is to provide a lightweight and flexible system through which users will be able to customise their own storage infrastructure.

The system consists of a minimal, simple set of fully customisable components, orthogonal in one another. In particular such a system aims to be flexible and portable in the sense that users will be able to bind to, update and manipulate data and programs without regard to:

- the location of the actors, the data or the programs. This means that a given program will work anywhere (with the appropriate infrastructure installed), regardless of its physical location or that of the data accessed.
- the storage technology employed.
- the network technology employed.

In such a system stores should not impose their own structure on the information stored (structure neutrality). This means that the same store may be used for information of various structures. In addition, the data stored and/or extracted from a store should be able to be interpreted in different ways.

Another requirement for XBase is that it should be able to treat different components of the system as data. As an example, it should be able for a store to be stored in another store. These components should be defined in a general way (interfaces) in order to allow programs to operate over different styles of stores in a uniform way. This means that programs should be able to access other programs and data in the same way no matter what style of store it is used – instantiate interfaces to abstract over concrete syntax of store.
Finally, data and programs should be stored in such a way that will allow historical views of data and programs.

# 2 The Rest of the World

Most of the existing systems provide a "heavyweight", complex approach in storing and retrieving data and programs. Each of the systems mentioned below provide a separate mechanism for storing data.
In addition, in order for these systems to become flexible and efficient, they provide more than it is needed and in most of the cases they are not optimised for particular applications.
Current paradigms for binding to data in programming languages are: file systems, programming languages, various storage systems, WWW, various internet applications.

## 2.1 File Systems

Various programming language/file system bindings, using appropriate sets of system calls:

- C - Unix style
- C - Windows style
- Java - Unix/Windows

File system implementation:

- inodes
- FAT etc

## 2.2 Storage

- relational database
- persistent store – binding to data; nature of stores/pointers themselves
- transactional store of various types
- XML document repository
- private data in public place e.g. on web page
- "objects" with shared state c.f. weather map example

## 2.3 Web

Conventional web servers and browsers.

Process involved when a hyper-link is clicked on:

html → key extracting interpreter → string → caster[key] → key → store → string → caster[html] → html

### 2.4 Other Internet

Various internet applications:

- CDDB
- napster
- gnutella
- freenet
- google

### 2.5 Data with Multiple Interpretations

- XML and programming language objects / historical views / bitmaps

## 3 Main Aspects of the XBase System (Meeting the Requirements)

The XBase system uses various features of the system/architectures described in section 2. There are two types of bindings: *keys to data* and *names to keys*. A possible general flow of information when saving data in a store is: *data => key =>name*. A possible general flow of information when retrieving data from a store is: *name=>key=>data*. Data can be retrieved by providing either a name or a key.

The main philosophy of the system is that data is stored in a uniform way without regard to the particular format (in some cases type). This means for example that an XML document can be stored in the same repository as a word document. However, the data retrieved from a store have to be interpreted before any further usage. More than one interpretation may be imposed over data. These interpretations are specified by an actor and may exist concurrently. The system is flexible enough to allow any user to create its own interpretation.

Another aspect of the system is related to protection. Actors may not be involved in the process of protecting data directly. This means that there is neither Unix style (e.g. <user>+go) nor type system protection. Instead data is protected by possibly requiring an actor to name the data or to interpret it (by using encryption/decryption algorithms) or by both.

Data can only be appended in a store, which means that there is no notion of erase. This implies that various versions of the same piece of data may exist in a store. An external mechanism may be built on top of that facility in order to allow the support of historical views.

All the aspects of the system described above are expressed in terms of component definitions. A particular set of interfaces is provided, each of which abstracts over a concrete definition of the corresponding component. These components are described in the next section.

# 4 Design

## 4.1 Components and Supported Interfaces

There are four main concepts defined in the XBase system. These are: ***Store***, ***Namer***, ***Caster***, ***Interpreter***. Other sub-components used by the system are ***Name***, ***Key***, ***Bitstring*** each of which represents a name, a key and data respectively.
The ***Store*** provides a one-one mapping between ***Keys*** and ***Bitstrings***. This mapping, which is under the control of the store, may fail if store fails, or if an invalid key is presented. The interface to a ***Store*** consists of three procedures, which are:

**put**: *BitString -> Key*
**get**: *Key -> BitString or error*
**getStoreID**: -> *BitString*

The ***Caster*** translates between entities and representations (***BitStrings***) stored in a ***Store***. It involves coercion and sometimes various computations and validity checking. The interface to a ***Caster*** consists of two procedures, which are (assuming that the entity's type is *t*):

**reflect**: *BitString -> t or error*
**reify**: *t -> BitString or error*

The ***Interpreter*** translates between BitStrings. It involves arbitrary computation such as encryption. The interface to an ***Interpreter*** consists of two procedures, which are:

**interpret**: *BitString -> BitString*

The ***Namer*** provides a many-many mapping between ***Names*** and ***Keys***. A ***Name*** may be bound to multiple ***Keys*** (set retrieval). A ***Key*** may be bound to multiple ***Names*** (aliasing). The mapping, which is under the control of caller, may be updated so given ***Name*** may refer to various ***Keys*** over time. The interface to a ***Namer*** consists of three procedures, which are:

**lookup**: *Name -> set[Key]*
**bind**: *Name,Key*
**unbind**: *Name,Key*

## 4.2 Using and Composing the XBase Components

The components defined in section 4.1 can be composed in various ways depending on the use case. An example use case of retrieving and putting from/in a store of an exam paper document is shown in Figure 1.

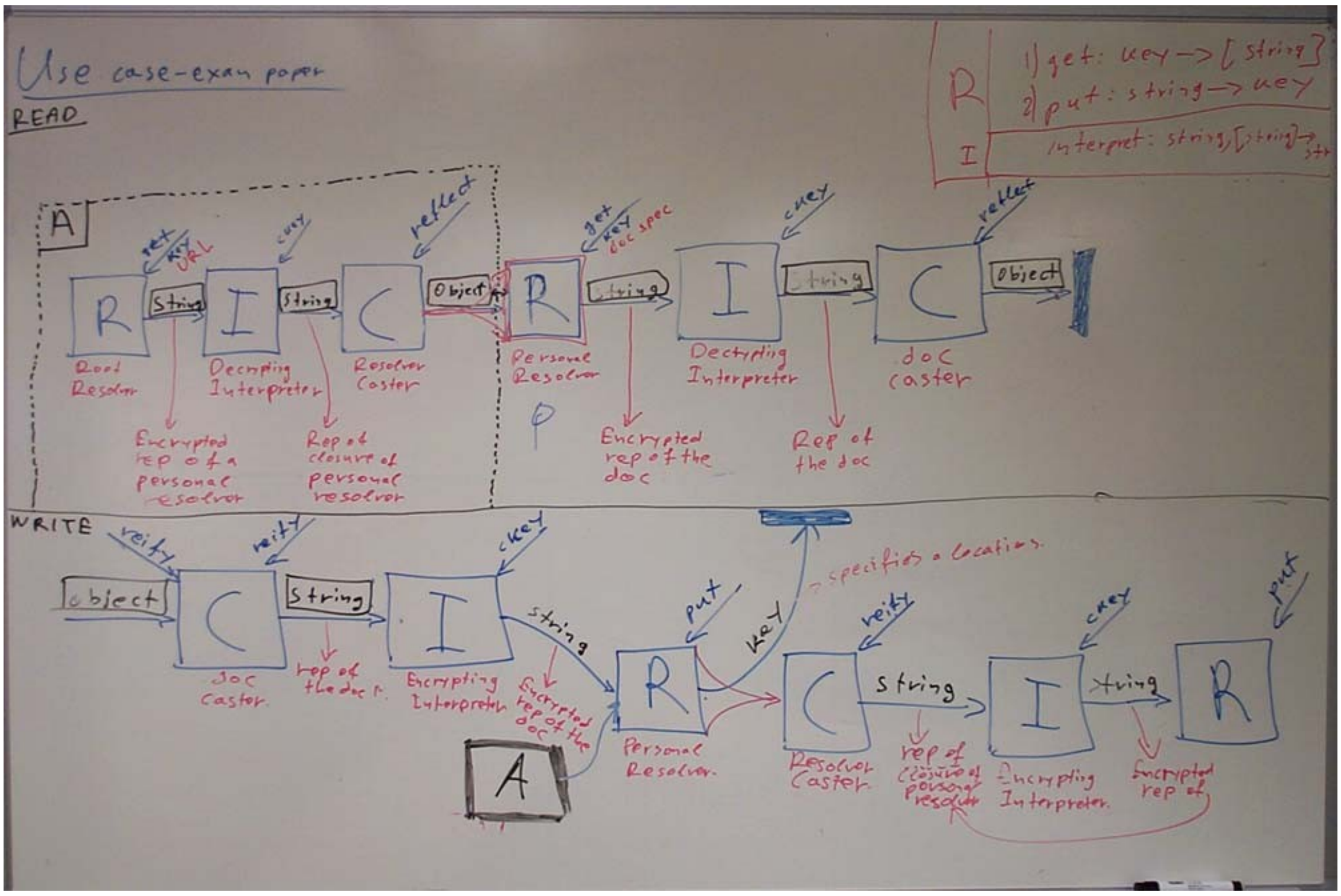

**Figure 1:** An example use case

In the case of reading a document, a personal store (PS) is retrieved from a root store (RS). The resulting representation of PS is decrypted (using a decrypting interpreter) and reflected (using a store caster). The resulting store entity is presented with the document's key, producing its representation. This representation is decrypted (using a document's interpreter) and reflected (using a document's caster) in order to produce the desired entity. A ***Namer*** could be used in order to resolve the document's key by presenting a requiring a name rather than a key.

In order to write a document the reverse process is performed as shown at the bottom of Figure 1.

The particular use case assumes that the components already exist before composition. However, since these components are data themselves, they can be retrieved from any other store in the same way as data. For example, a representation of a caster can be retrieved from a store, which is then possibly interpreted (using a decrypting algorithm) and reflected to a caster entity (using a caster for that particular caster.

## 4.3 Styles and Dimensions

XBase provides the model upon which a user can build a particular implementation of any component. Different use cases require different styles of ***Stores***, ***Casters***, ***Interpreters*** and ***Namers*** been generated. The only requirement for writing a component is that it has to conform to the corresponding interface. In this document we provide some example implementations of various components. The dimensions for each of the component are described in the following sub-sections.

### 4.3.1 Stores

Different styles of Stores are categorised according to the following dimensions:

- Locality: that is where the information is stored. Example stores are:
    - a local store: that is an instantiation of a store which exists in the same place with the program that manipulates it (at the implementation level there may be variations such as):
        - using a single file for backing storage of the bindings,
        - using multiple files for backing storage of the bindings
        - use no backing storage for the bindings at all but keep them transient.
    - a proxy store: that is a store that knows about other stores (this may be recursive, i.e. that store knows about another store that knows about other stores …),
- Shareability: that is whether a store can be “exported” to the rest of the world. An example is a shareable store, which can be either a local or a proxy store and access is allowed by any other store. A store been non-shareable means that access is granted only to the owner,
- Persistence: an example is a root store which is probably the only fixed component in the system. It can be either a local store or a proxy store and it is the analogous to the root of persistence in persistent systems,

### 4.3.2 Casters

Different styles of Casters are categorised according to the following dimensions:

- Generality: that is how general is the type of the entity about to be flattened (e.g. a single type, a more complex type).
- Flattening Strategy: that is what is the particular decision made for flattening the entity (e.g. should the representation be self-contained or not).

Example casters are:

- XBase component casters: that is casters that produce or flatten XBase components. An example of such a caster is a store caster.
- Intermediate form casters: that is casters which can be applied on a group of application specific casters. An example of such a caster is a Java serialisation caster.
- Application specific casters: that is casters which comply with particular format of entities/representations. An example of such a caster is an XML document caster.

### 4.3.3 Interpreters

Different styles of Interpreters are categorised according to the following dimensions:

- Transformation policy: that is what do we want to achieve with the transformation (e.g. encryption, compression).

### 4.3.4 Namers

Different styles of Namers are categorised according to the following dimensions:

- Locality: that is how and where do we save the bindings. Examples of such Namers are:
    - A Namer with backing storage: it saves the *name, key* bindings in a file in a local host.
    - A Namer with no backing storage: it uses no backing storage at all. *Name, key* binding are transient, and the only way to make that namer persistent is to reify it, using the appropriate caster, and put it in a store

as data. Retrieval of the namer involves reflecting the previously stored representation.

- Persistence: that is whether we want to make the bindings persistent or not.

A concrete implementation of all the ideas mentioned above has already been implemented and is described in the next section.

# 5 Implementation

This section describes a set of Java interfaces which implement the basic ideas of the XBase model as well as various Java classes that implement those interfaces. The current implementation has been structured into two Java packages containing the classes shown in fig 2.

Package *xbase.model* contains interfaces and some concrete implementations of XBase components. Package *xbase.exceptions* contains exceptions that are thrown when a failure occurs. This exception is caught either by the system or is required to be caught by the user.

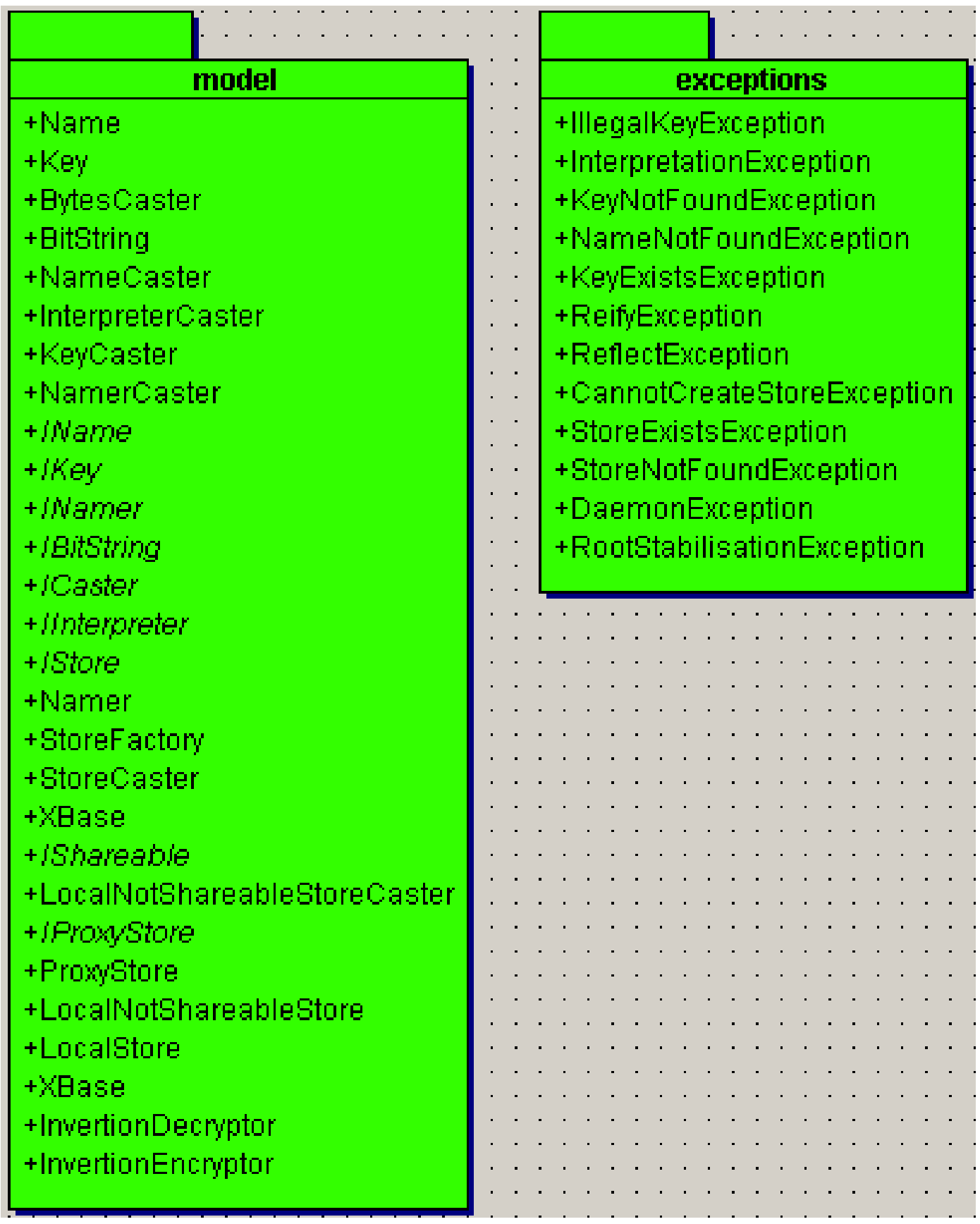

**Figure 2:** Packages in the current implementation

In the sections that follow, the interfaces are coloured in dark blue background and their concrete implementations as well as other classes in cyan. Note that the blue lines denote associations and the green dashed lines denote inheritance.

## 5.1 The Store

As mentioned earlier different styles of stores are defined in the XBase system. These are shown in Figure 3. All types of Stores inherit for interface *IStore*, which means that implementation of *put*, *get* and *getStoreID* operations should be provided. In the figure, two classes inherit directly from that interface (*LocalNotShareableStore* and *LocalStore*). Although both are styles of a local store what makes them different is the particular implementation of the backing storage. The former uses a single file to store data. However, the latter uses a file per data unit, that is a file per ***BitString***, and supports multiple backing storage locations details of which are specified by instances of class *BackingStorageDetails*.
A special style of an *IStore* is defined by interface *IProxyStore*, all implementations of which must specify operations *addTarget*, *removeTarget* and *lookupTarget* as well as the operations inherited from *IStore* interface (such an implementation is provided through class *ProxyStore*). *IProxyStore* defines a store that knows about other stores. A target can be either another store (instantiation of the *IStore* interface) or a URL specifying the location of another store.

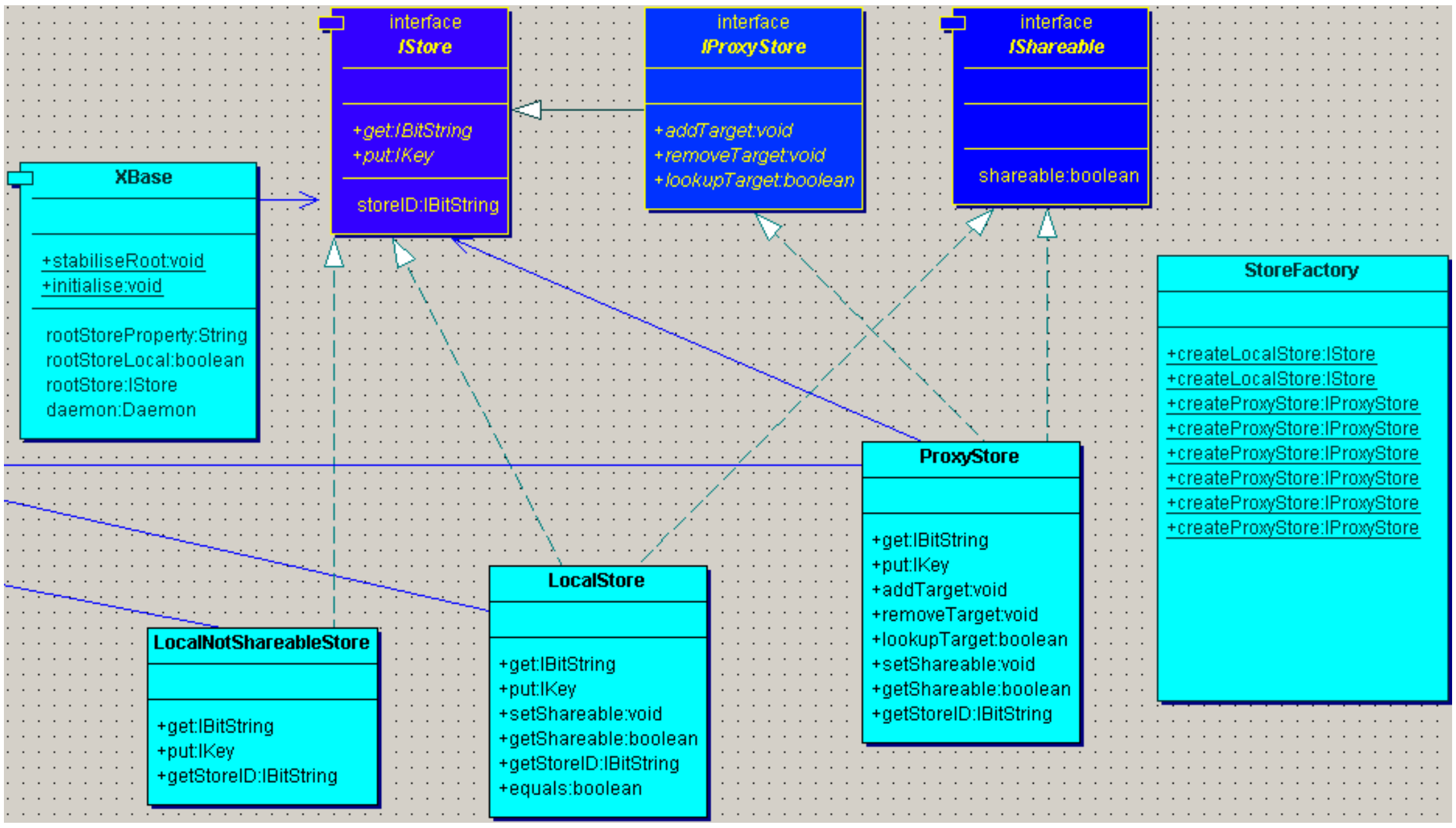

**Figure 3:** Different styles of Stores defined in the XBase system.

Any style of store may be accessible by other stores only if it is shareable. Shareability is defined by having a store inherit from interface *IShareable*. An actor may switch shareability on/off. If shareability is off this means that store is private and cannot be accessible by any other store. Shareability of a store may even be changed remotely through an *IProxyStore* as well as locally through a program. In order to achieve the former, a unique id is used.
Switching the shareability of a store to on means that the particular store is accessible by any other store. Behind the scenes the store registers itself as shareable in a daemon that starts automatically either at system start-up or when switching the shareability of the store. The daemon accepts various requests from other stores.
Creating new instances of a store from within a program can only be achieved through a store factory. As shown in Figure 3, all the styles of stores mentioned above may be created by invoking the appropriate static method of class *StoreFactory*.

A **Root Store** may be retrieved through the static method *XBase.getRootStore* in a way that is hidden from an actor. However, it is not a special type of store, since it implements the *IStore* interface. It can also implement the *IProxyStore* interface in which case the way other stores or data are retrieved is done differently from having a local store as root.
Root stores can be stabilised by invoking method *XBase.stabiliseRoot* (see later section). Stabilisation means that a store is reified in order to produce a representation which is saved in a particular place in the local node. In order to create the ***Root Store*** on system restart, that representation is retrieved and reflected to an instantiation of *IStore*.

## 5.2 The Caster

Each *XBase* entity (including components in the model) requires its corresponding Caster in order to provide the facility to flatten the entity into a general enough representation and recreate the entity from that representation. All the casters inherit from interface *ICaster*, which means that they must implement methods *reflect* and *reify*.
Fig 4 shows various casters defined as part of the XBase model. Class *BytesCaster* is an example of an **Intermediate Form Caster** introduced earlier. Classes *NameCaster*, *KeyCaster*, *StoreCaster*, *NamerCaster*, *InterpreterCaster* are examples of **XBase Component Casters**. Classes *WordDocumentCaster*, *AudioDocumentCaster*, *XMLCaster*, shown in Figure 17, are examples of **Application Specific Casters**.

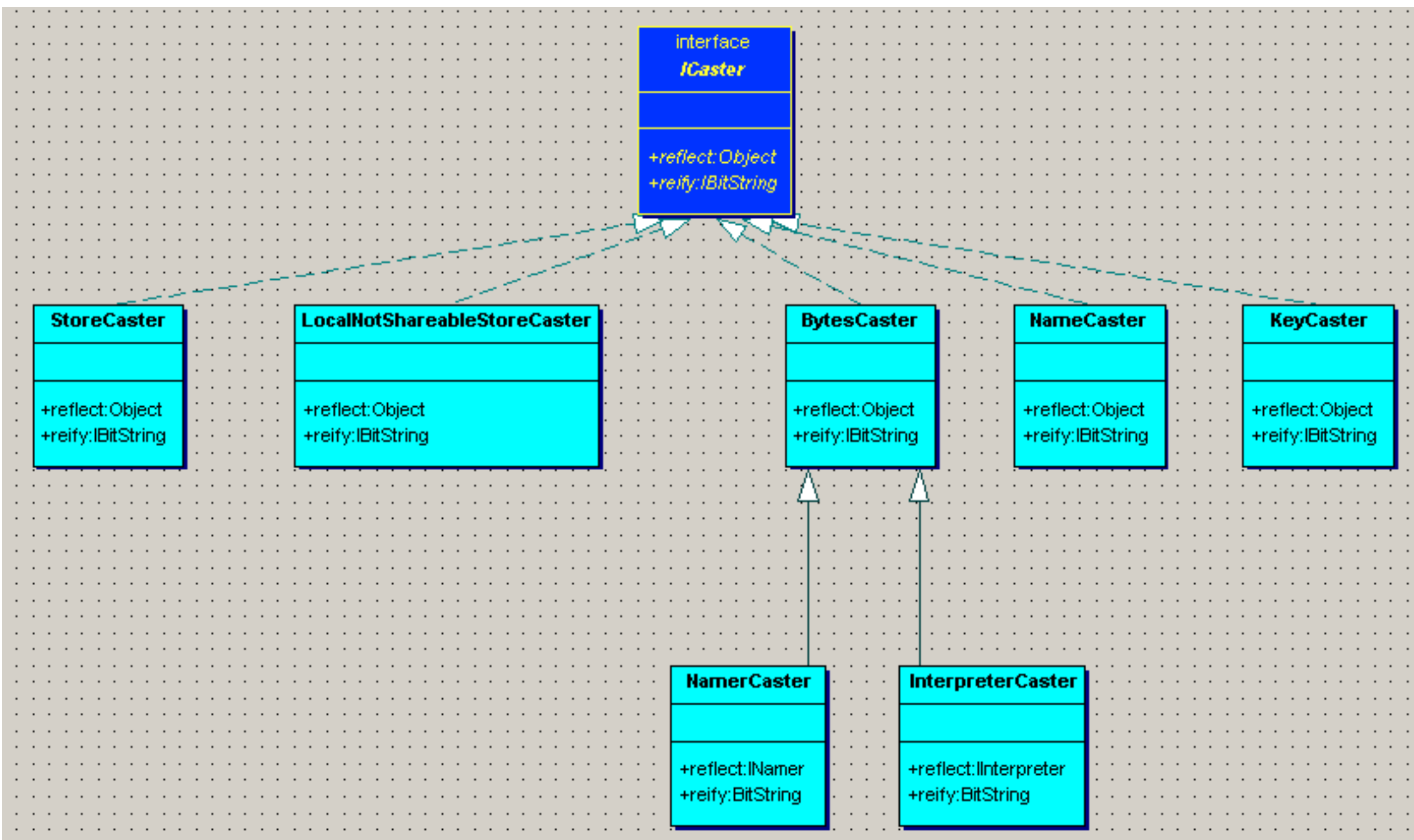

**Figure 4:** Different styles of Casters as part of the XBase model

An interesting case of a caster is the *StoreCaster*, which provides the procedural interface to flatten a store as well as create a store from a representation. The particular caster uses DOM to flatten a store into an XML representation and create a store from an XML document. An example of such a representation is shown in

Figure 5. The XML document represents an *IProxyStore* that knows about a local store, another *IProxyStore* and other *IProxyStore* on different nodes. The local store has two backing storage locations. Each backing storage location consists of a url and a list of those files that contained in the url at the time of reification.

```xml
<?xml version="1.0" encoding="UTF-8"?>
<network ID="2" SHAREABLE="true">
  <localStores>
    <local ID="1" SHAREABLE="true">
      <backingStorage ID="1">
        <url>file:///Z:/Projects/XBase/prototype/DefaultStore0</url>
        <files>
          <file ID="1">7da32c717e8b34e9126489e221ccf194</file>
          <file ID="2">24679927be100dc8d6df280d88f20d77</file>
        </files>
      </backingStorage>
      <backingStorage ID="2">
        <url>file:///Z:/Projects/XBase/prototype/DefaultStore1</url>
        <files>
          <file ID="1">7da32c717e8b34e9126489e221ccf194</file>
        </files>
      </backingStorage>
    </local>
    <network ID="3" SHAREABLE="true">
      <localStores/>
      <remoteNodes>
        <url>http://tsipouro.dcs.st-and.ac.uk:17000</url>
      </remoteNodes>
    </network>
  </localStores>
  <remoteNodes>
    <url>http://ouzo.dcs.st-and.ac.uk:17000</url>
    <url>http://burgie.dcs.st-and.ac.uk:17000</url>
    <url>http://panda.dcs.st-and.ac.uk:17000</url>
  </remoteNodes>
</network>
```

**Figure 5:** An example XML representation of a store

### 5.3 Other Components

Other components that are part of the XBase model are defined by the interfaces and classes shown in fig 6. Concrete implementations of ***Interpreters*** and ***Namers*** are provided by instantiating the *IInterpreter* and the *INamer* interface respectively. The former requires implementation of operation *interpret*, which translates between representations. The latter requires the implementation of operations *lookup*, *bind* and *unbind*, which manipulate name, key bindings. An example ***Namer*** is class *Namer*, which stores the bindings in a Java *Hashtable*. Example interpreters are provided through classes *InvertionEncryptor* and *InvertionDecryptor*, which reverse the order of bytes in the given representation.
Names, keys and representations are defined in the model through interfaces *IName*, *IKey* and *IBitString* respectively. Concrete implementations of those interfaces are classes *Name*, *Key* and *BitString*, which effectively wrap and manipulate a string, a string and an array of bytes respectively.

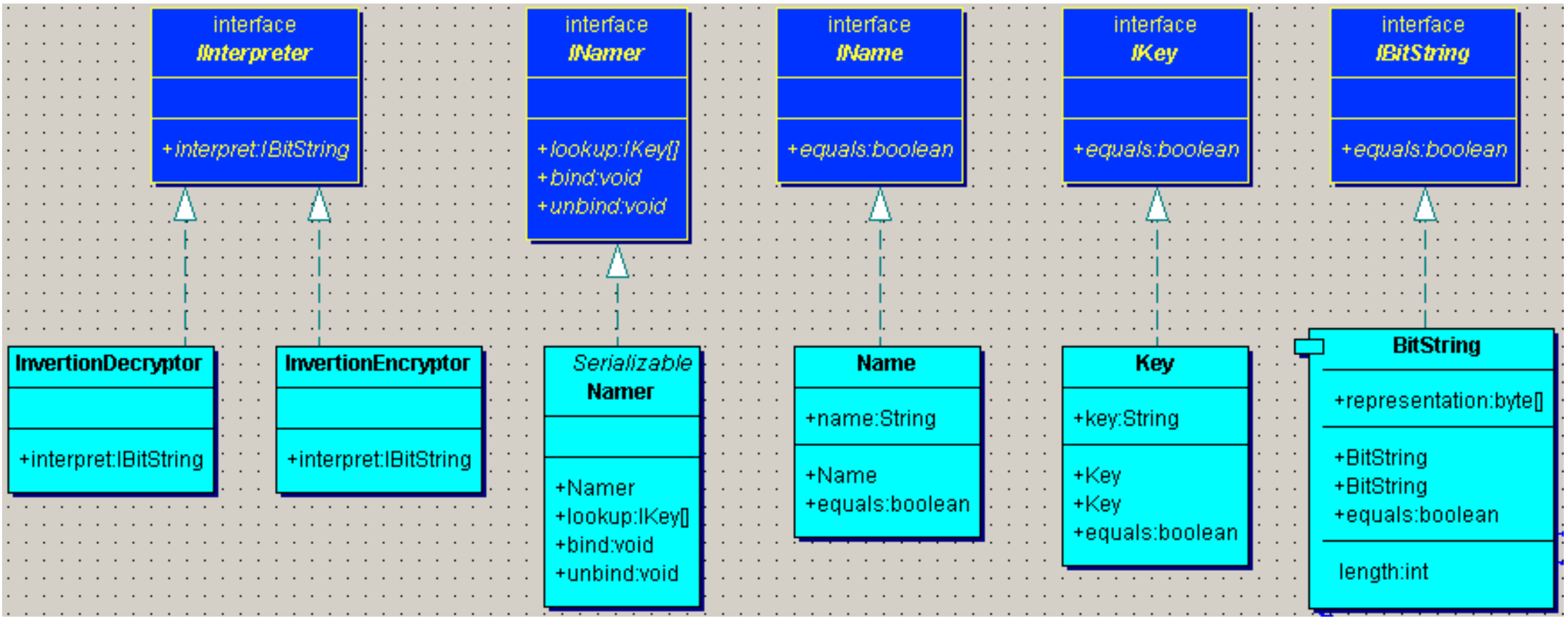

**Figure 6:** Various XBase components used in the model

## 6 Various Use Cases

In this section we provide particular examples of composing the XBase components described above. The use cases presented here demonstrate:

- root store usage (Use Case 1),
- root store usage with other XBase components (Use Case 2),
- store operations over different types of entities (Use Case 3),
- local store usage (Use Case 4),
- proxy store as a root store (Use Case 5),
- store operations over a remote node (Use Case 6),
- re-directing store operations from one remote node to another (Use Case 7).

The sections that follow contain descriptions of various scenarios and their corresponding Java source code that shows the way that XBase components may be composed to satisfy the scenario.

### 6.1 Use Case 1 (UC1)

The aim of UC1 is to demonstrate the usage of a root store. The root store is a local store. The scenario is as follows:

> "*On node **tsipouro** create an instance of class Person, add it in the root store, and retrieve it using the pre-recorded key.*"

An abstract diagrammatic description of the use case is shown in Figure 7.

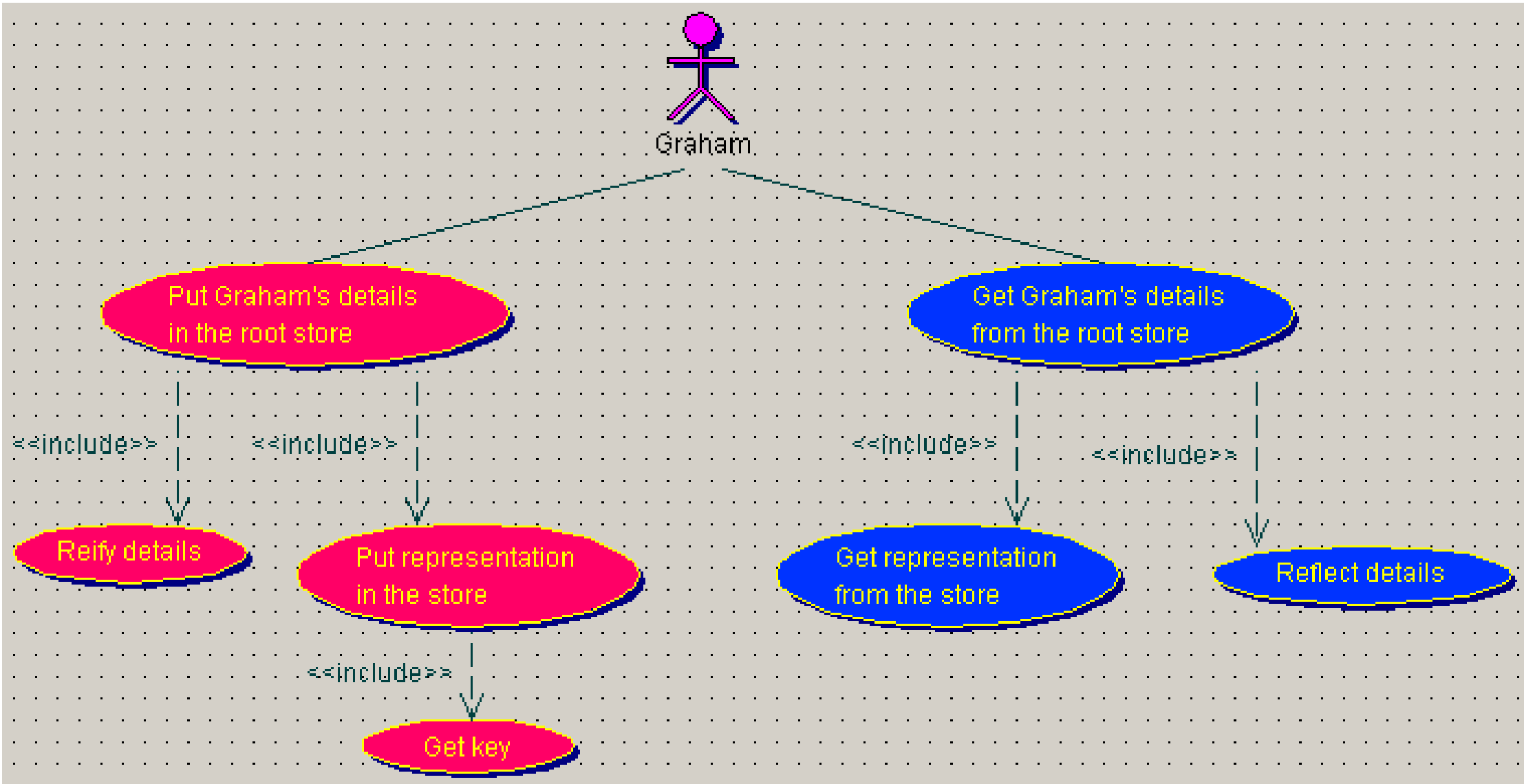

**Figure 7:** Use case diagram for UC1

The classes used for satisfying the needs of UC1 are shown in Figure 8. Class *Person* is accompanied by its corresponding caster (class *PersonCaster*). The representation of a *Person* instance is a string containing the name and the age.

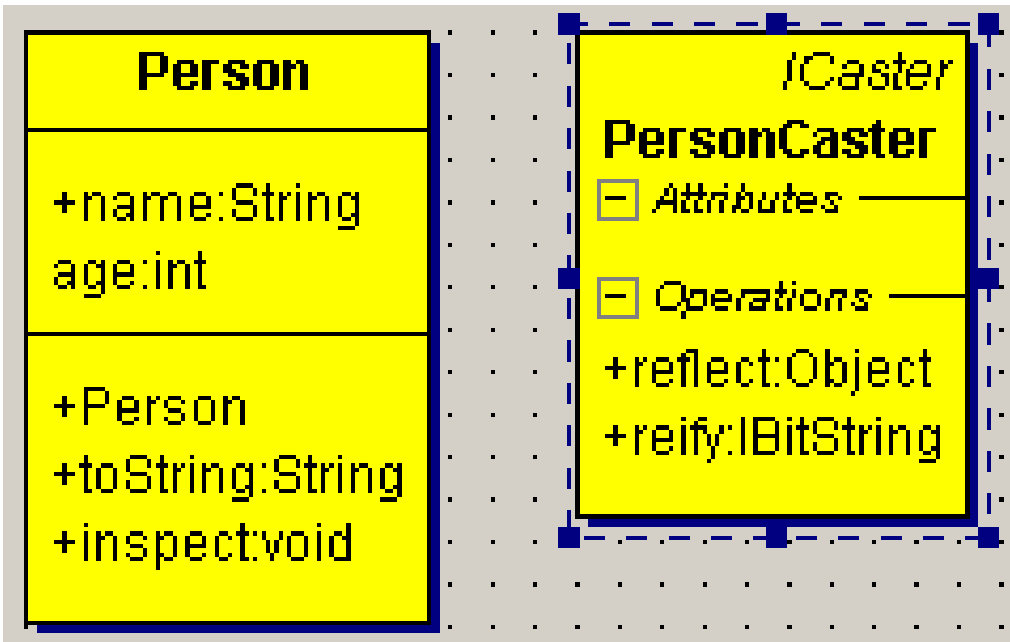

**Figure 8:** Classes used in UC1

The source for achieving the storing part of UC1, by making use of interpreters and namer as well, in shown in Figure 9.

```
// Create an instance of class Person
Person graham = new Person("Graham", 35);

// Retrieve an instance of the root store
IStore rootStore = XBase.getRootStore();

// Create a Caster specifically written for persons
PersonCaster personCaster = new PersonCaster();

// Flatten the person into a bitstring
IBitString personRep = personCaster.reify(graham);

// Put the representation in the root store
IKey grahamKey = rootStore.put(personRep);
```

**Figure 9:** Storing part of UC1

The source code for retrieving a person with a name previously recorded, is shown in Figure 10.

```
// Retrieve an instance of the root store
IStore rootStoreAgain = XBase.getRootStore();

// Retrieve the representation of the person using the key
IBitString grahamRep = rootStoreAgain.get( grahamKey  );

// Create a Caster specifically written for persons
PersonCaster personCaster = new PersonCaster();

// Recreate the object using the representation
Person reflectedGraham = (Person)personCaster.reflect( grahamRep );

// Print out the details
reflectedGraham.inspect();
```

**Figure 10:** Retrieval part of UC1

The result of executing the code illustrated above is shown in Figure 11.

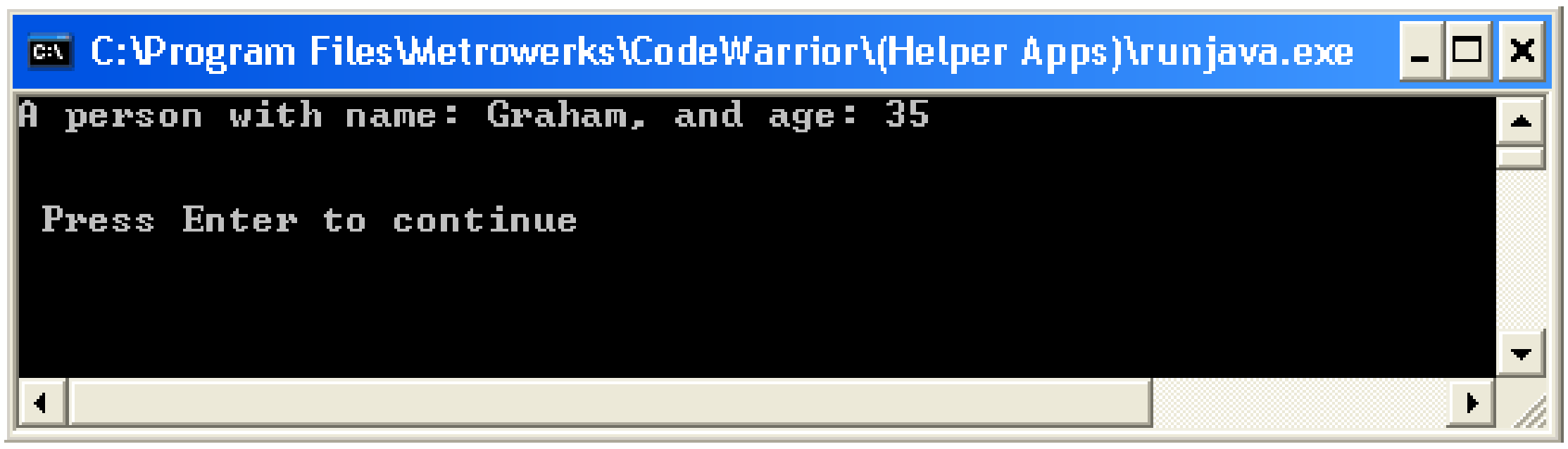

**Figure 11:** Result of UC1

## 6.2 Use Case 2 (UC2)

The aim of UC2 is to demonstrate the usage of a root store together with other XBase components, such as a namer. UC2 extends UC1, which means that the root store is still a local store. The scenario is as follows:

> "*On node* ***tsipouro*** *create an instance of class Person, encrypt it, add it in the root store and record the name. Using that name retrieve it from the root store, decrypt it and display its details.*"

An abstract diagrammatic description of the use case is shown in Figure 12.

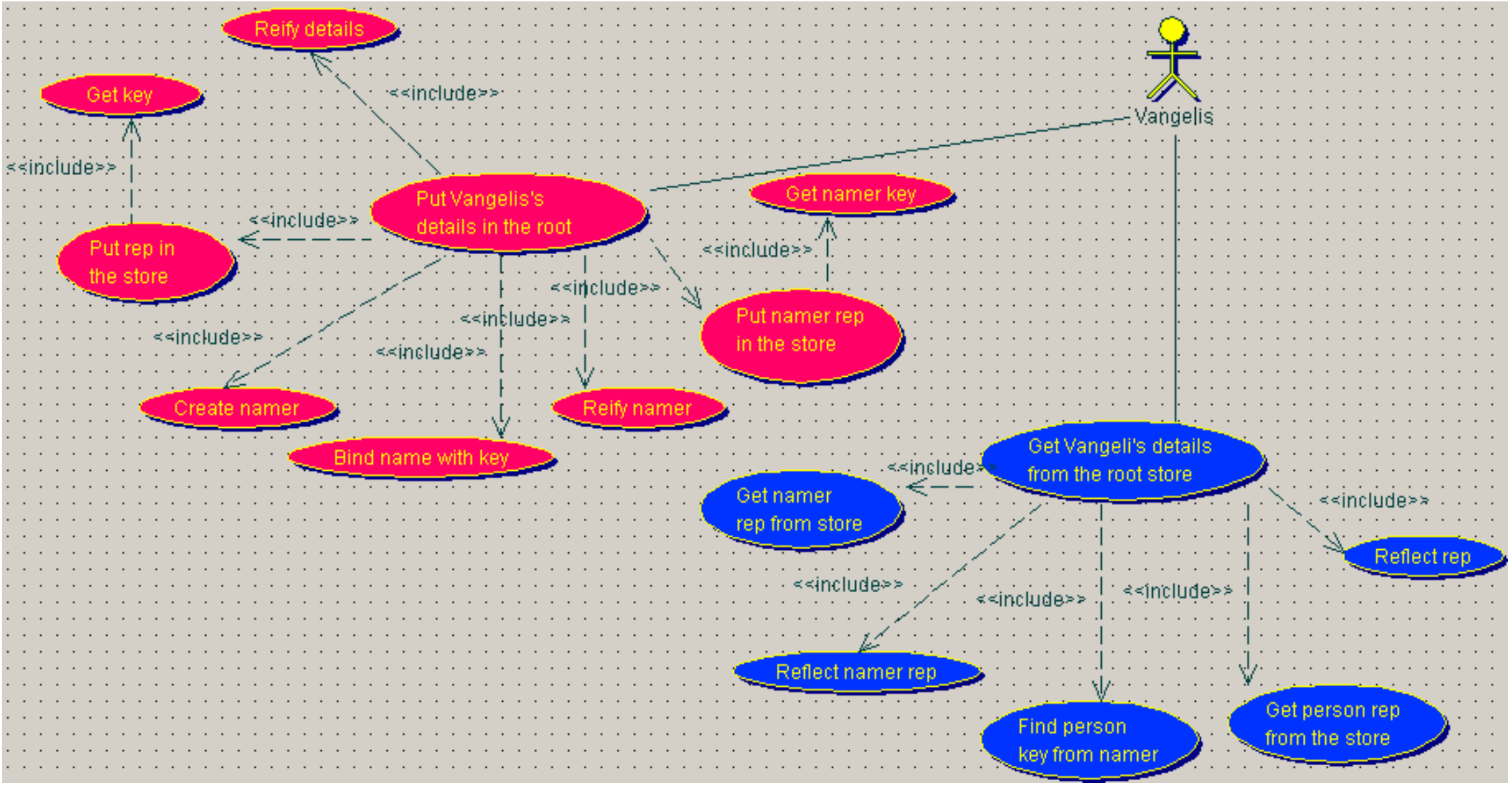

**Figure 12:** Use case diagram for UC2

The classes used for satisfying the needs of UC2 are shown in Figure 8. Class *Person* is accompanied by its corresponding caster (class *PersonCaster*). The representation of a *Person* instance is a string containing the name and the age.
The source for achieving the storing part of UC2, by making use of a namer as well, in shown in Figure 13.

```
// Create an instance of class Person
Person vangelis = new Person("Vangelis", 29);

// Create a globally accessible namer
myNamer = new Namer();

// Retrieve an instance of the root store
IStore rootStore = XBase.getRootStore();

// Create a Caster specifically written for persons
PersonCaster personCaster = new PersonCaster();

// Flatten the person into a bitstring
IBitString personRep = personCaster.reify( vangelis );

// Create an encryption interpreter
IInterpreter encryptionInterpreter = new InvertionEncryptor();

// Encrypt the representation of person
IBitString encryptedPersonRep = encryptionInterpreter.interpret(personRep);

// Put the encrypted representation in the root store
IKey vangelisKey = rootStore.put(encryptedPersonRep);
```

```
// Bind the resulted key with a name
IName vangelisName = new Name(aPerson.name);
INamer myNamer = new Namer();
myNamer.bind( vangelisName, vangelisKey );

// Create a caster specifically written for namers.
ICaster namerCaster = new NamerCaster();

// Flatten the namer into a bitstring
IBitString namerRep = namerCaster.reify(myNamer);

// Put the namer rep in the store
IKey namerKey = rootStore.put(namerRep);
```

**Figure 13:** Storing part of UC2

The source code for retrieving a person with a name and a key for a namer recorded previously, is shown in Figure 14.

```
// Retrieve an instance of the root store
IStore rootStoreAgain = XBase.getRootStore();

// Retrieve the namer rep from the store using the pre-recorded key
IBitString namerRep = rootStoreAgain.get(namerKey);

// Create a caster specifically written for namers.
ICaster namerCaster = new NamerCaster();

// Create a namer from a bitstring
INamer reflectedNamer = (INamer)namerCaster.reflect(namerRep);

// Find the key using the person's name
IKey[] vangelisKeys = reflectedNamer.lookup( new Name("Vangelis") );

// Retrieve the representation of the person using the key
IBitString vangelisRep = rootStoreAgain.get( vangelisKeys[0] );

// Create an decryption interpreter
IInterpreter decryptionInterpreter = new InvertionDecryptor();

// Decrypt the representation of person
IBitString decryptedPersonRep = decryptionInterpreter.interpret( vangelisRep );

// Create a Caster specifically written for persons
PersonCaster personCaster = new PersonCaster();

// Recreate the object using the representation
Person reflectedVangelis = (Person)personCaster.reflect(decryptedPersonRep);

// Print out the details
reflectedVangelis.inspect();
```

**Figure 14:** Retrieval part of UC2

The result of executing the code illustrated above is shown in Figure 15.

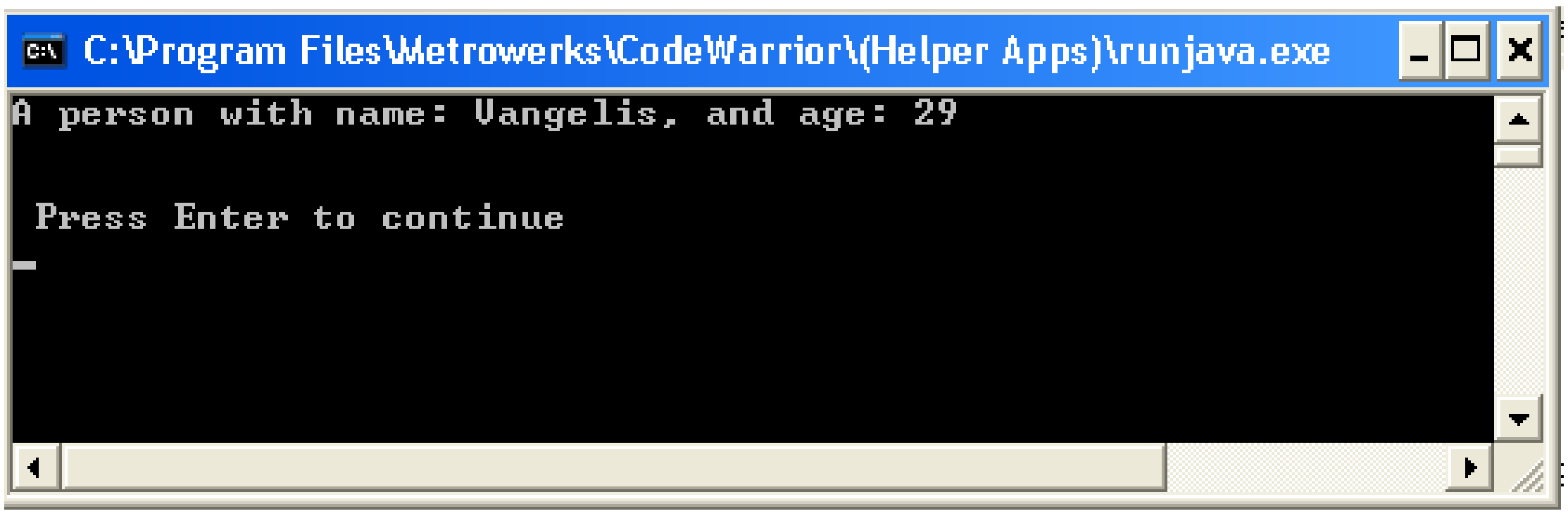

**Figure 15:** Result of UC2

## 6.3 Use Case 3 (UC3)

The aim of UC3 is to demonstrate store operations over different type of entities. Example entities used here are: a word document, an audio clip, an image and an xml document. The scenario is as follows:

> "*On node* ***tsipouro*** *add representations of a word document, an audio clip, an image and an xml document in the root store. Retrieve those entities from the root store and show the results.*"

An abstract diagrammatic description of the use case is shown in Figure 16.

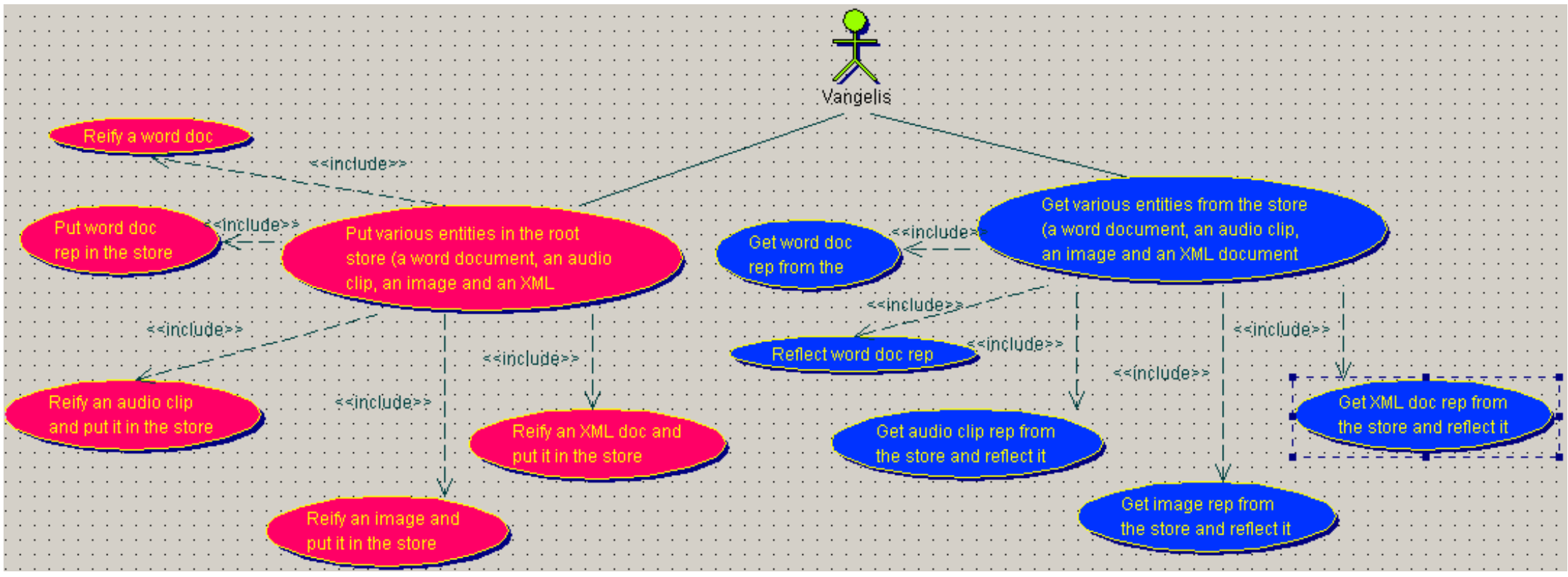

**Figure 16:** Use case diagram for UC3

The classes used for satisfying the needs of UC3 are shown in Figure 16. The different entities are represented by classes *WordDocument*, *AudioClip*, *ImageIcon* (part of standard JDK) and *Document* (part of the Xerces DOM implementation). Each entity is accompanied by its corresponding caster (classes *WordDocumentCaster*, *AudioClipCaster*, *ImageCaster* and *XMLCaster* respectively). Note that for word documents and audio clips, the representation encapsulated is an array of bytes.

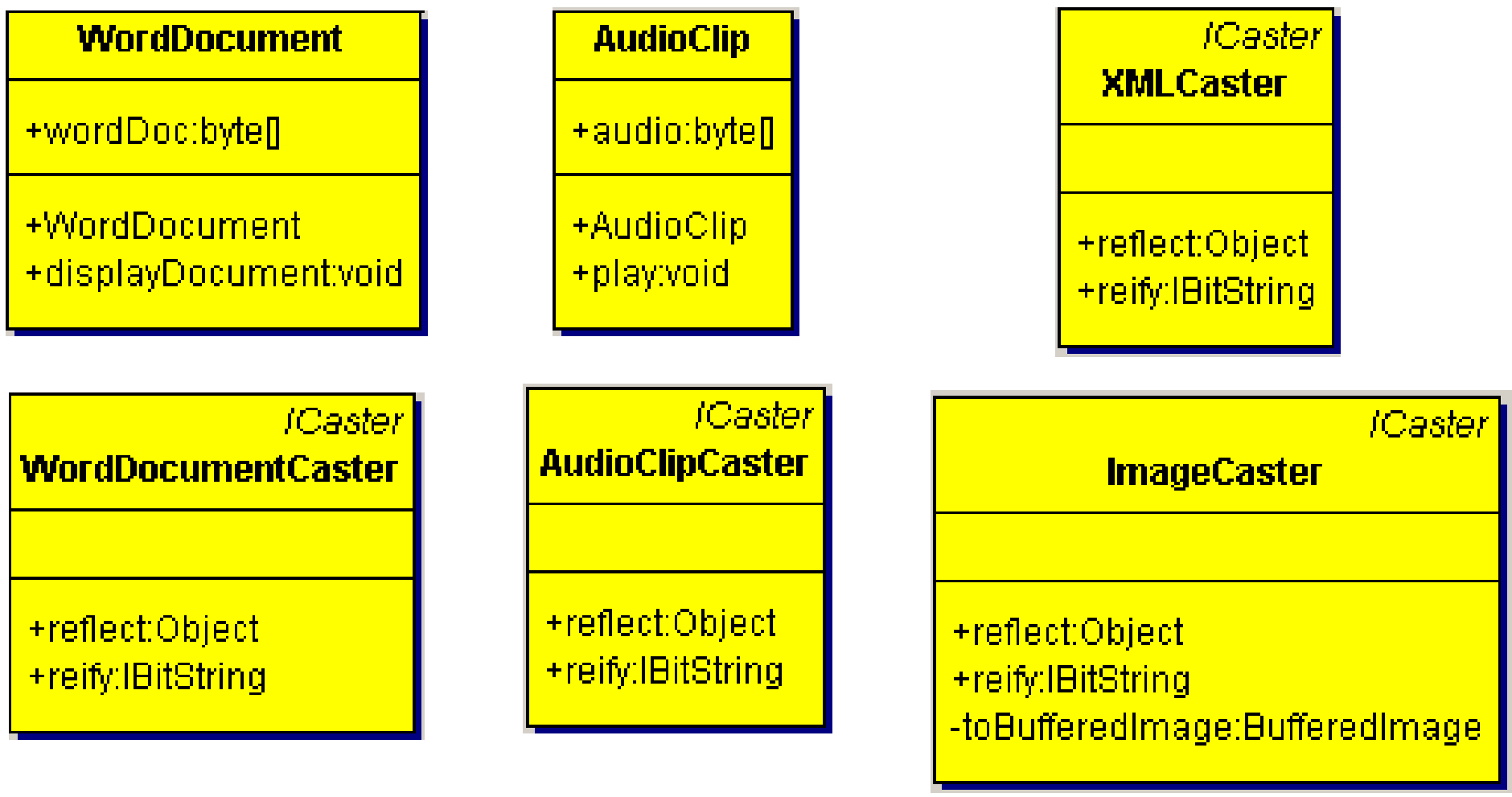

**Figure 17:** Classes used in UC3

The source code for achieving the storing part of the scenario is shown in Figure 18. Note that method *Util.readBytesFromAFile* reads the bytes stored in the given file. Also method *Util.createDOMFromFile* parses an XML document stored in the given file.

```
// *********** Create objects ***********
// Create a word document object
WordDocument wordDoc = new WordDocument( Util.readBytesFromAFile("testWord.doc") );

// Create an audio clip object
AudioClip audioClip = new AudioClip( Util.readBytesFromAFile("testaudio.wav") );

// Create an image object
ImageIcon vangelisImage = new ImageIcon("vangelis.gif");

// Create another image object
ImageIcon grahamImage = new ImageIcon("graham.gif");

// Create a DOM document object from a string representation of an XML document
Document xmlDoc = Util.createDOMFromFile("examplesDir/xbaseMembers.xml");

// ************** Put entities in the root store and record the keys *********
// Retrieve an instance of the root store
IStore rootStore = XBase.getRootStore();

// Create a caster specifically written for word documents
ICaster wordDocCaster = new WordDocumentCaster();

// Flatten the word doc into a bitstring
IBitString wordDocRep = wordDocCaster.reify( wordDoc );

// Put the representation of the word doc in the root store
IKey wordDocKey = rootStore.put( wordDocRep );

// Create a caster specifically written for audio clips
ICaster audioClipCaster = new AudioClipCaster();

// Flatten the audio clip into a bitstring
IBitString audioClipRep = audioClipCaster.reify( audioClip );

// Put the representation of the audio clip in the root store
```

```
IKey audioClipKey = rootStore.put( audioClipRep );

// Create a caster specifically written for images
ICaster imageCaster = new ImageCaster();

// Flatten the image into a bitstring
IBitString imageRep = imageCaster.reify( image );

// Put the representation of the image in the root store
IKey imageKey = rootStore.put( imageRep );

// Create a caster specifically written for XML documents
ICaster xmlDocCaster = new XMLCaster();

// Flatten the XML document into a bitstring
IBitString xmlDocRep = xmlDocCaster.reify( xmlDoc );

// Put the representation of the XML document in the root store
IKey xmlDocKey = rootStore.put( xmlDocRep );
```

**Figure 18:** The storing part of UC3

The source code for retrieving the entities using their pre-recorded keys is shown in Figure 19.

```
// Retrieve an instance of the root store
IStore rootStore = XBase.getRootStore();

// ***** Retrieve the entities from the root store using the pre-recorded keys *****

// Retrieve the word doc from the store
IBitString wordDocRep = rootStore.get( wordDocKey );

// Create a caster specifically written for word documents
ICaster wordDocCaster = new WordDocumentCaster();

// Create a word doc object from the bitstring
WordDocument wordDoc = (WordDocument)wordDocCaster.reflect( wordDocRep );

// Retrieve the audio clip from the store
IBitString audioClipRep = rootStore.get( audioClipKey );

// Create a caster specifically written for audio clips
ICaster audioClipCaster = new AudioClipCaster();

// Create an audio clip object from the bitstring
AudioClip audioClip = (AudioClip)audioClipCaster.reflect( audioClipRep );

// Get the representation of the image
IBitString imageRep = rootStore.get( imageKey );

// Create a caster specifically written for Images
ICaster imageCaster = new ImageCaster();

// Create the image object using the caster over the representation
ImageIcon reflectedImage = (ImageIcon)imageCaster.reflect( imageRep );

// Retrieve the XML document from the store.
```

```
IBitString xmlDocRep = rootStore.get( xmlDocKey );

// Create a caster specifically written for XML documents
ICaster xmlDocCaster = new XMLCaster();

// Create an XML document DOM object from the bitstring
Document reflectedXMLDoc = (Document)xmlDocCaster.reflect( xmlDocRep );

// ******** Manipulate the entities appropriately ****************
// Display the word document
reflectedWordDoc.displayDocument();

// Play the audio clip
reflectedAudioClip.play();

// Display the images
Util.displayImage( reflectedVangelisImage, "Vangelis" );
Util.displayImage( reflectedGrahamImage, "Graham" );

// Display the xml document
Util.displayXMLDocument( reflectedXMLDoc );
```

**Figure 19:** The retrieval part of UC3

The result of executing the code illustrated above displays the word document, plays the audio clip, displays the images and displays the XML DOM tree. This is shown in Figure 20 (word document), Figure 21 (images) and Figure 22 (XML document).

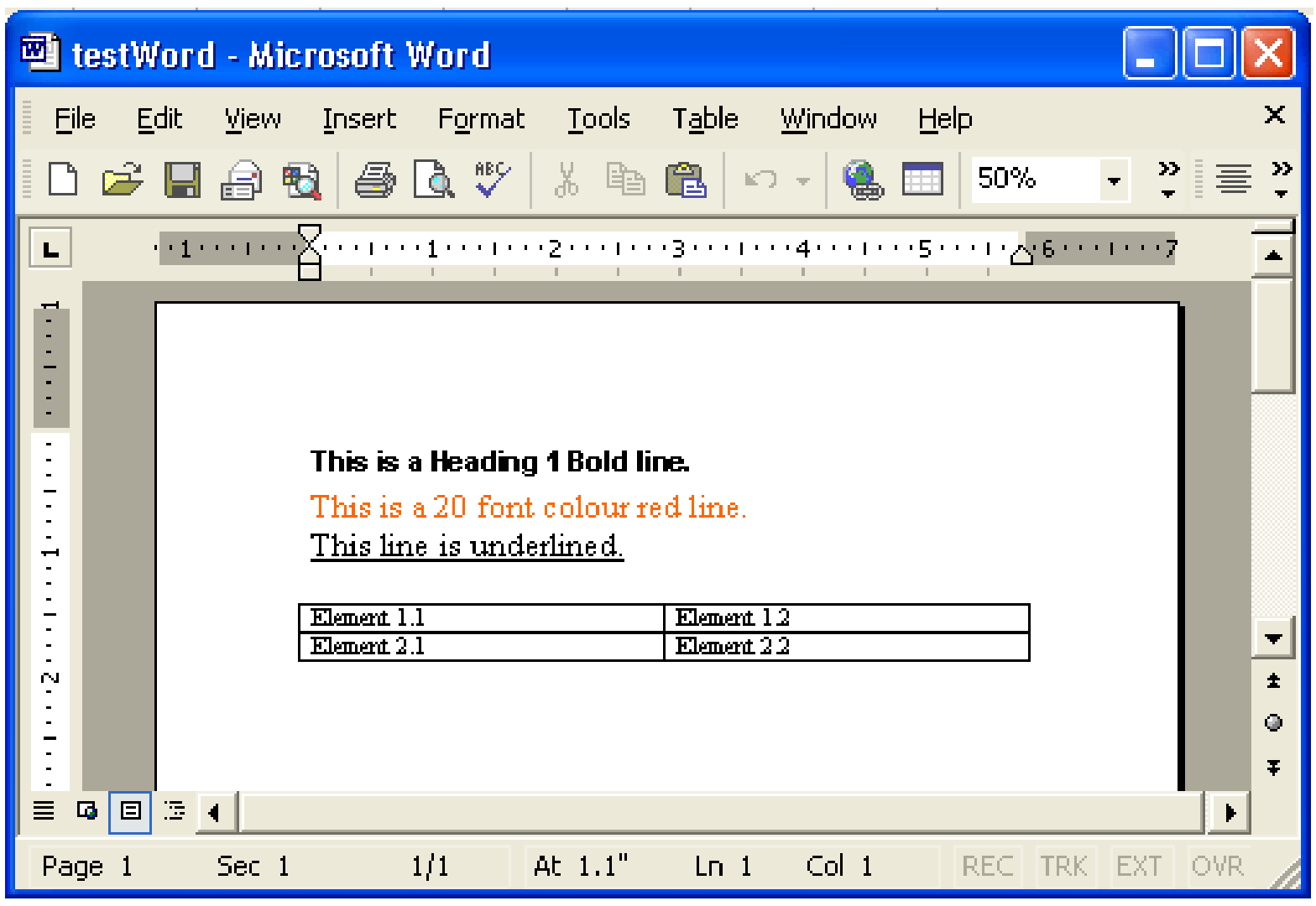

**Figure 20:** Result of UC3 (Word Document)

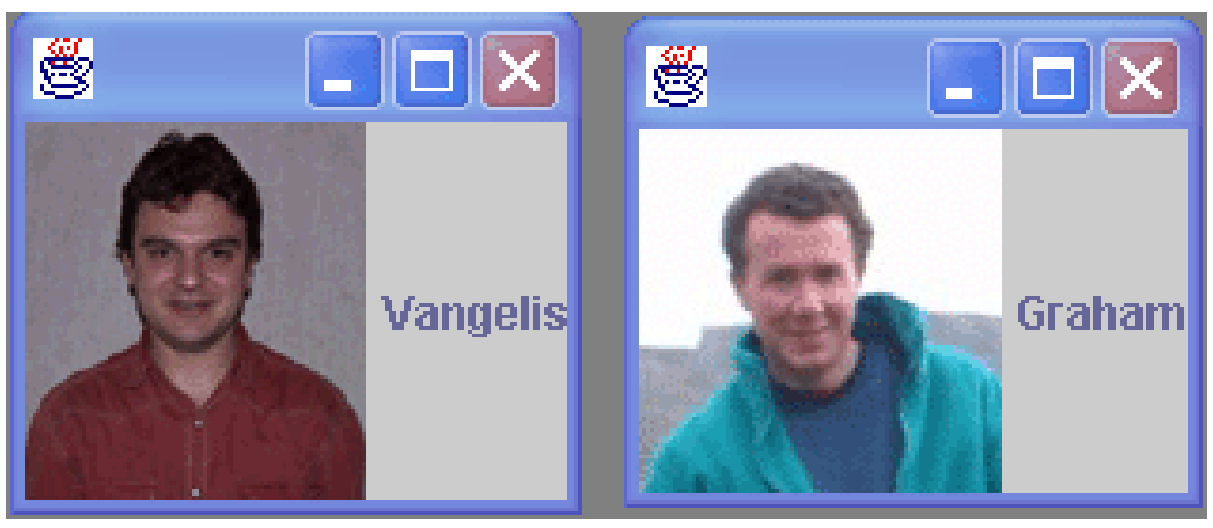

**Figure 21:** Result of UC3 (Image)

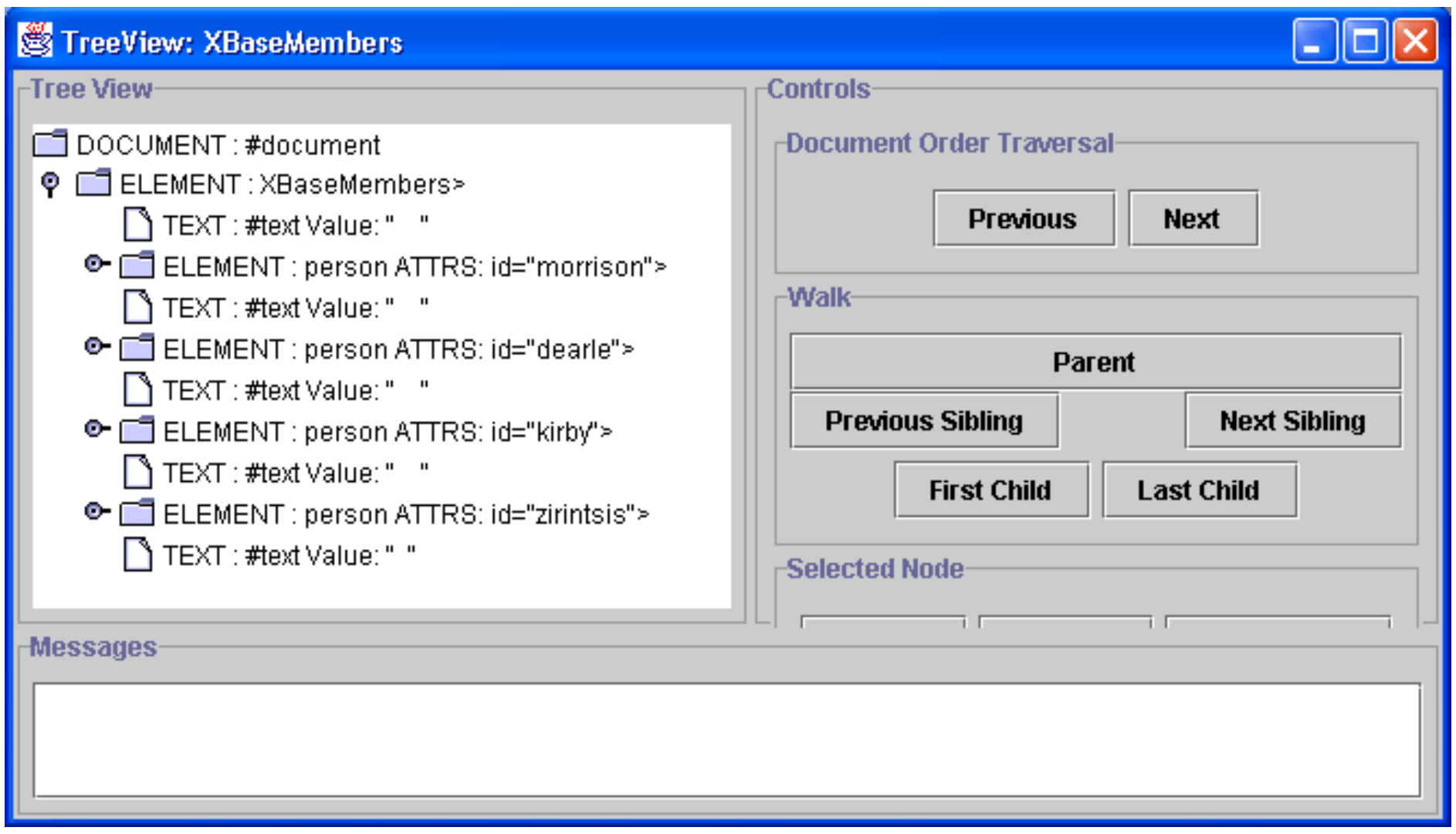

**Figure 22:** Result of UC3 (XML DOM tree)

## 6.4 Use Case 4 (UC4)

The aim of UC3 is to demonstrate the usage of local stores when these are treated as any other XBase entity, that is been reified, added in another local store and reflected. The scenario is as follows:

> "*On node **tsipouro** create a local store and add a representation of an XML document. Make that local store persistent by reifying it and adding it in a root store. Retrieve the XML document from the root store through that local store and display the tree.*"

An abstract diagrammatic description of the use case is shown in Figure 23.

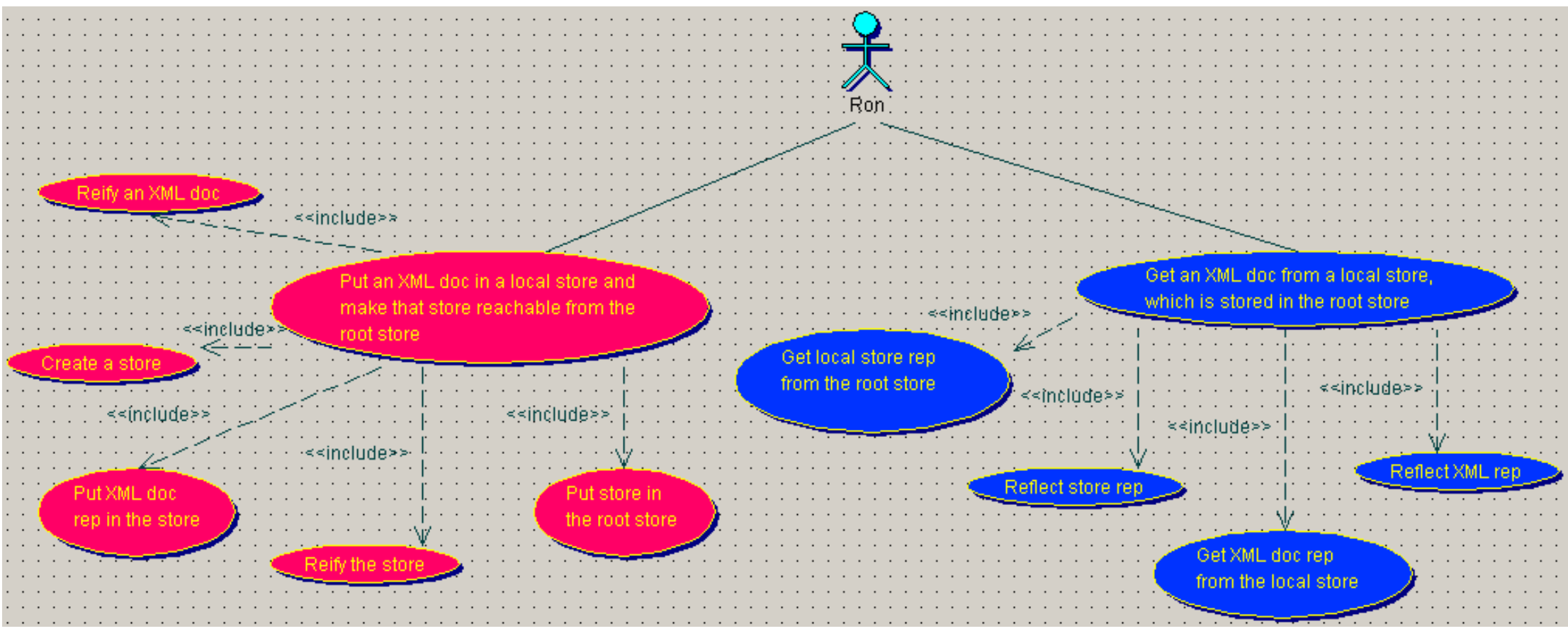

**Figure 23:** Use case diagram for UC4

In order to satisfy the needs of UC4 a caster is required to reflect and reify an XML DOM object. As shown in Figure 17, class *XMLCaster* provides an implementation of such a caster.

The source code for achieving the storing part of UC4 is shown in Figure 27.

```
// Create a DOM document object from a string representation of an XML document
Document xmlDoc = Util.createDOMFromFile("xbaseMembers.xml");

// Retrieve an instance of the root store
IStore rootStore = XBase.getRootStore();

// Create a caster specifically written for XML documents
ICaster xmlDocCaster = new XMLCaster();

// Flatten the XML document into a bitstring
IBitString xmlDocRep = xmlDocCaster.reify( xmlDoc );

// Create a local store
IStore myLocalStore = StoreFactory.createLocalStore();

// Put the representation of the XML document in the local store
IKey xmlDocKey = myLocalStore.put( xmlDocRep );

// Create a caster specifically written for stores
ICaster storeCaster = new StoreCaster();

// Reify the store
IBitString storeRep = storeCaster.reify( myLocalStore );

// Put the representation of the local store in the root store
IKey storeKey = rootStore.put( storeRep );
```

**Figure 24:** The storing part of UC4

The source code for achieving the retrieval part of UC4 is shown in

```
// Retrieve an instance of the root store
IStore rootStore = XBase.getRootStore();

// Retrieve a representation of the local store
IBitString storeRep = rootStore.get(  storeKey  );

// Create a caster specifically written for stores
ICaster storeCaster = new StoreCaster();

// Create a store object using the appropriate caster
IStore myLocalStore = (IStore)storeCaster.reflect( storeRep );

// Retrieve the XML document from the store.
IBitString xmlDocRep = myLocalStore.get( xmlDocKey  );

// Create a caster specifically written for XML documents
ICaster xmlDocCaster = new XMLCaster();

// Create an XML document DOM object from the bitstring
Document reflectedXMLDoc = (Document)xmlDocCaster.reflect( xmlDocRep );

// Display the xml document
Util.displayXMLDocument( reflectedXMLDoc );
```

**Figure 25:** Retrieval part of UC4

Executing the code illustrated above results in displaying the XML DOM tree shown in Figure 22.

## 6.5 Use Case 5 (UC5)

The aim of UC5 is to demonstrate the usage of a root store been a proxy store. The scenario is as follows:

> "*On node **tsipouro** create a shareable local store, add an XML document and make that local store reachable from the root store (root store is a proxy store). Finally, retrieve the XML document and display the XML DOM tree.*"

An abstract diagrammatic description of the use case is shown in Figure 26.

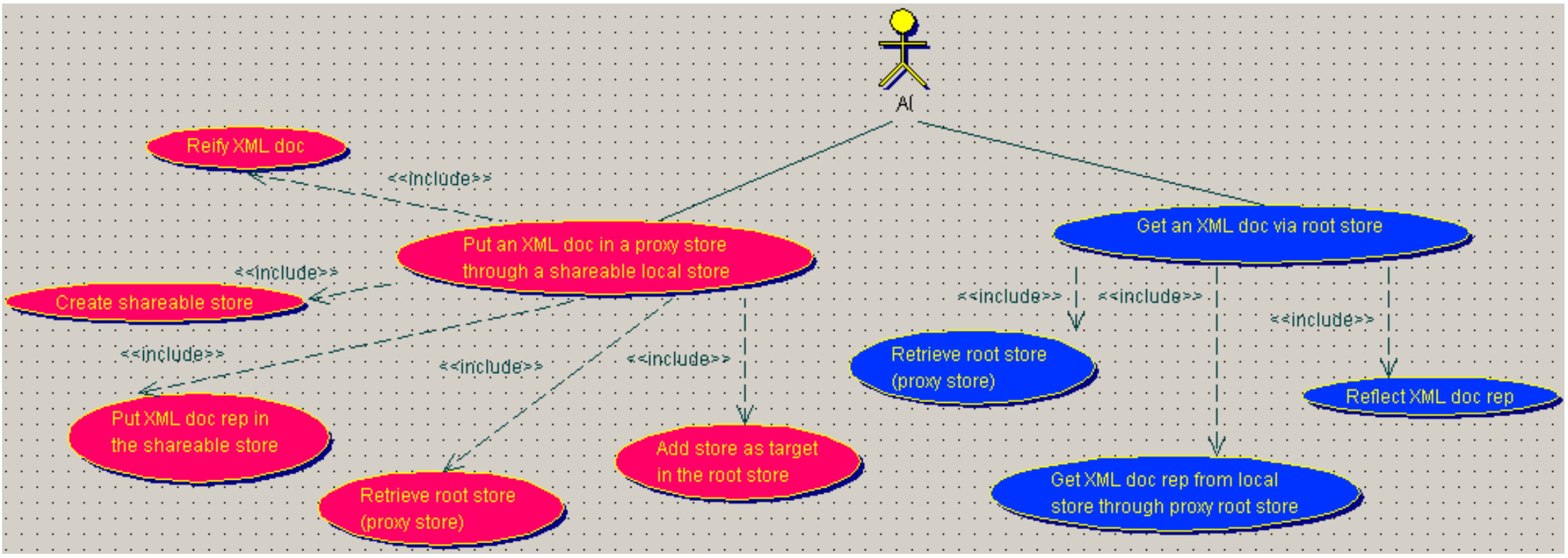

**Figure 26:** Use case diagram for UC5

In order to satisfy the needs of UC5 a caster is required to reflect and reify an XML DOM object. As shown in Figure 17, class *XMLCaster* provides an implementation of such a caster.

The source code for achieving the storing part of UC5 is shown in Figure 27.

```
// Create a DOM document object from a string representation of an XML document
Document xmlDoc = Util.createDOMFromFile("xbaseMembers.xml");

// Set the type of the root store to be a proxy store
XBase.setRootStoreProperty("proxy");

// Retrieve an instance of the root store.
// We expect to have an instance of IProxyStore, that is why we perform a check first.
// If it is any other type of store, do nothing
if (XBase.getRootStore() instanceof IProxyStore) { // It is an instance of IProxyStore

    // Retrieve an instance of the root store
    IProxyStore rootStore = (IProxyStore)XBase.getRootStore();

    // Create a caster specifically written for XML documents
    ICaster xmlDocCaster = new XMLCaster();

    // Flatten the XML document into a bitstring
    IBitString xmlDocRep = xmlDocCaster.reify( xmlDoc );

    // Create a shareable local store
    IStore myLocalStore = StoreFactory.createLocalStore(true);

    // Put the representation of the XML document in that local store
    IKey xmlDocKey = myLocalStore.put( xmlDocRep );
```

```
        // Make the local stors reachable from the root store (root store is a proxy store)
        rootStore.addTarget(myLocalStore);

    } else throw new CannotCreateStoreException("Illegal type of store. IProxyStore expected.");
```

**Figure 27:** The storing part of UC5

The source code for achieving the retrieval part of UC5 is shown in Figure 28.

```
// Retrieve an instance of the root store.
// No check is performed because we are using the root store as an IStore and not as an IProxyStore.
IStore rootStore = XBase.getRootStore();

// Retrieve the XML document through the root store.
// Root store is a proxy store that knows about a local store
// Operation ROOTSTORE.get scans sequentially the store that is aware of.
// The only reachable store is shareable and performing a get returns the requested entity.
IBitString xmlDocRep = rootStore.get( xmlDocKey );

// Create a caster specifically written for XML documents
ICaster xmlDocCaster = new XMLCaster();

// Create an XML document DOM object from the bitstring
Document reflectedXMLDoc = (Document)xmlDocCaster.reflect( xmlDocRep );

// Display the xml document
Util.displayXMLDocument( reflectedXMLDoc );
```

**Figure 28:** Retrieval part of UC5

Note that by default stores are not shareable when they created by *StoreFactory* unless otherwise specified, as in this particular use case.

The resulting XML DOM tree is the same as the one shown in Figure 22. Figure 29 illustrates messages produced during the process of looking in stores reachable from the root. The first store found (store with ID *bd…953*) referenced through the proxy root store is shareable and contains the desired XML representation.

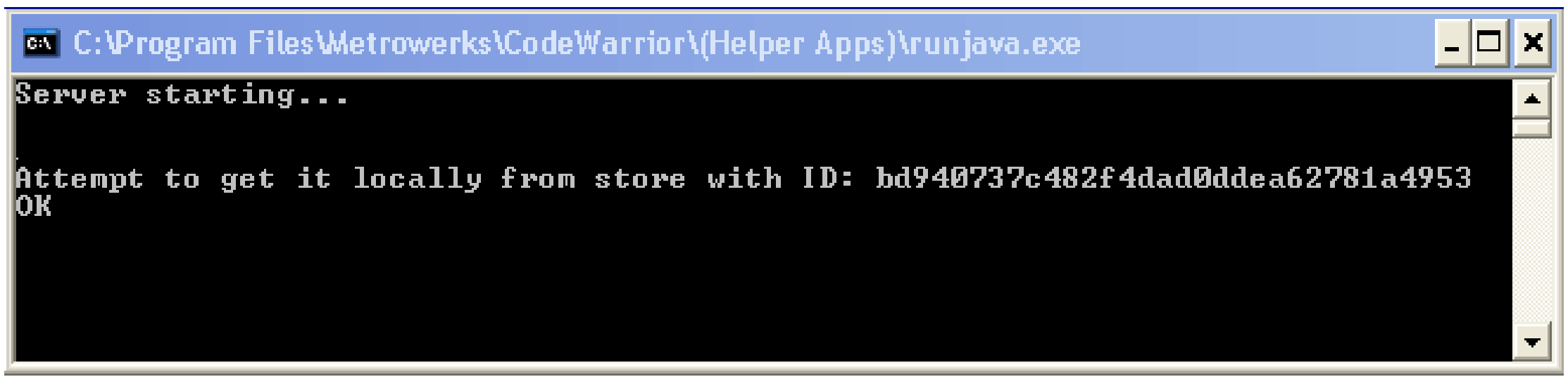

**Figure 29:** Console window messages when performing UC5

## 6.6 Use Case 6 (UC6)

The aim of UC6 is to demonstrate operations performed on stores in remote hosts. The scenario is as follows:

> "*On node **ouzo** create a shareable local store and add an Image. On node **tsipouro** create a proxy store that knows about the local stores* *on **ouzo** and add another image through that proxy store. Finally, retrieve both of the images from node **tsipouro**.*"

An abstract diagrammatic description of the use case is shown in Figure 30.

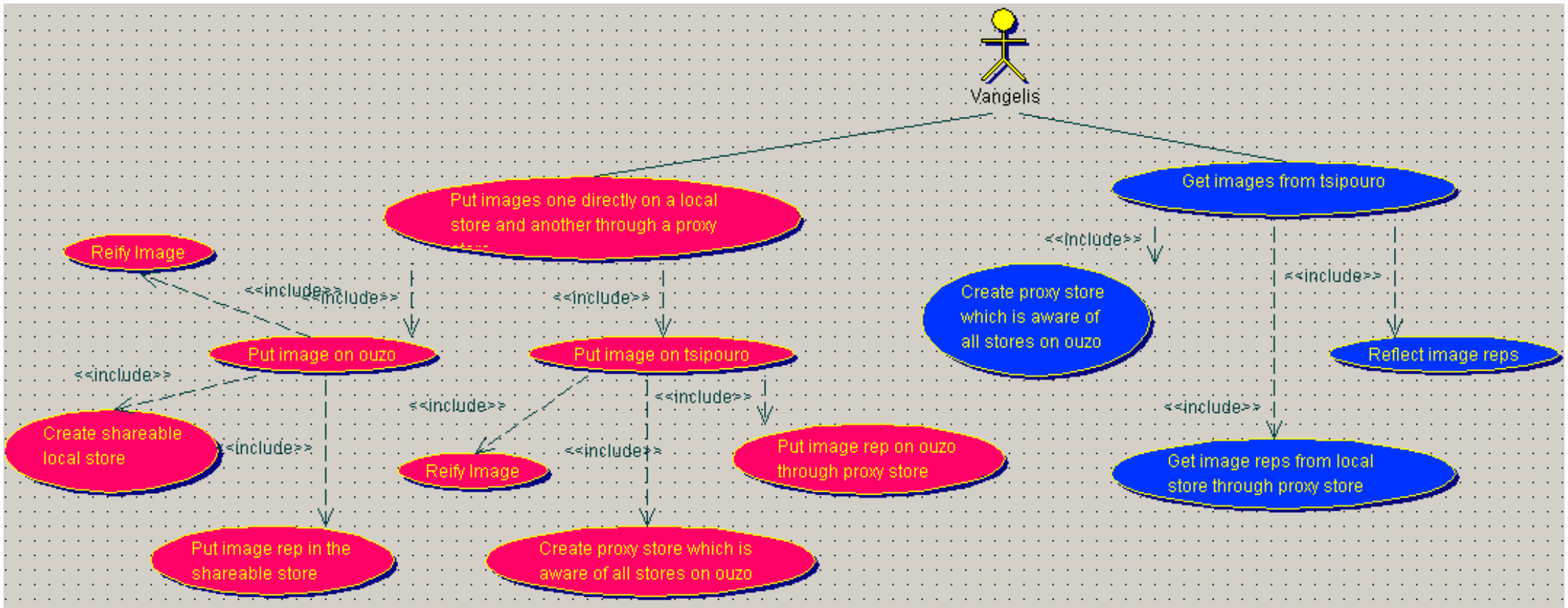

**Figure 30:** Use case diagram for UC6

In order to satisfy the needs of UC6 a caster is required to reflect and reify an image object. As shown in Figure 17, class *ImageCaster* provides an implementation of such a caster.
The source code for achieving the storing part of UC6 on **ouzo** is shown in Figure 31.

```
// Create an image object
ImageIcon vangelisImage = new ImageIcon("vangelis.gif");

// Create a caster specifically written for images
ICaster imageCaster = new ImageCaster();

// Flatten the image into a bitstring
IBitString imageRep = imageCaster.reify( image );

// Create a shareable local store
IStore myLocalStore = StoreFactory.createLocalStore(true);

// Put the representation of the image in the local store
IKey vangelisImageKey = myLocalStore.put( imageRep );
```

**Figure 31:** Storing part on **ouzo** of UC6

The source code for achieving the storing part of UC6 on **tsipouro** is shown in Figure 32. Note that method *Util.createURL* creates a URL Java object using the given host.

```
// Create an image object
ImageIcon grahamImage = new ImageIcon("graham.gif");

// Create a caster specifically written for images
ICaster imageCaster = new ImageCaster();

// Flatten the image into a bitstring
IBitString imageRep = imageCaster.reify( grahamImage );

// Create a proxy store that knows about host ouzo.
// The URL does not specify a particular store, which means that this proxy store will be aware of all
the shareable stores in the remote host.
```

```
IStore aProxyStore = StoreFactory.createProxyStore( Util.createURL("ouzo.dcs.st-and.ac.uk") );

// Put the encrypted representation of the image in the proxy store
// Behind the scenes:
// since the proxy store is not aware of any local stores, it will redirect the request to the available stores on remote hosts.
// The only available store on a remote host is on ouzo, and this is where the operation is performed.
IKey vangelis = aProxyStore.put( imageRep );
```

**Figure 32:** Storing part on **tsipouro** of UC6

The source code for achieving the retrieving part of UC6 on **tsipouro** is shown in Figure 33. Note that both of the images are retrieved using the pre-recorded keys.

```
// Create a proxy store that knows about node ouzo
// The URL does not specify a particular store, which means that this proxy store will be aware of all the shareable stores in the remote host.
IStore aProxyStore = StoreFactory.createProxyStore( Util.createURL("ouzo.dcs.st-and.ac.uk") );

// Get the representation of the first image
// Behind the scenes:
// since the proxy store is not aware of any local stores, it will redirect the request to the available stores on remote hosts.
// The only available store on a remote host is on ouzo, and this is where the operation is performed.
IBitString vangelisImageRep = aProxyStore.get( vangelisImageKey );

// Get the representation of the second image.
// The steps that are followed behind the scenes are the same as when retrieving the first image.
IBitString grahamImageRep = aProxyStore.get( grahamImageKey );

// Create a caster specifically written for Images
ICaster imageCaster = new ImageCaster();

// Create the image objects using the caster over the decrypted representation
ImageIcon reflectedVangelisImage = (ImageIcon)imageCaster.reflect( vangelisImageRep );
ImageIcon reflectedGrahamImage = (ImageIcon)imageCaster.reflect( grahamImageRep );

// Display the images
displayImage(reflectedVangelisImage );
displayImage(reflectedGrahamImage );
```

**Figure 33:** The retrieval part on **tsipouro** of UC6

The result of executing the code illustrated above is shown in Figure 21. Figure 34 illustrates messages produced during the process of retrieving the images Since, the proxy store on node **tsipouro** only references a store on node **ouzo** both of the requests are redirected to the latter. These requests are successful and both the representations of the images are sent back to **tsipouro** and displayed there.

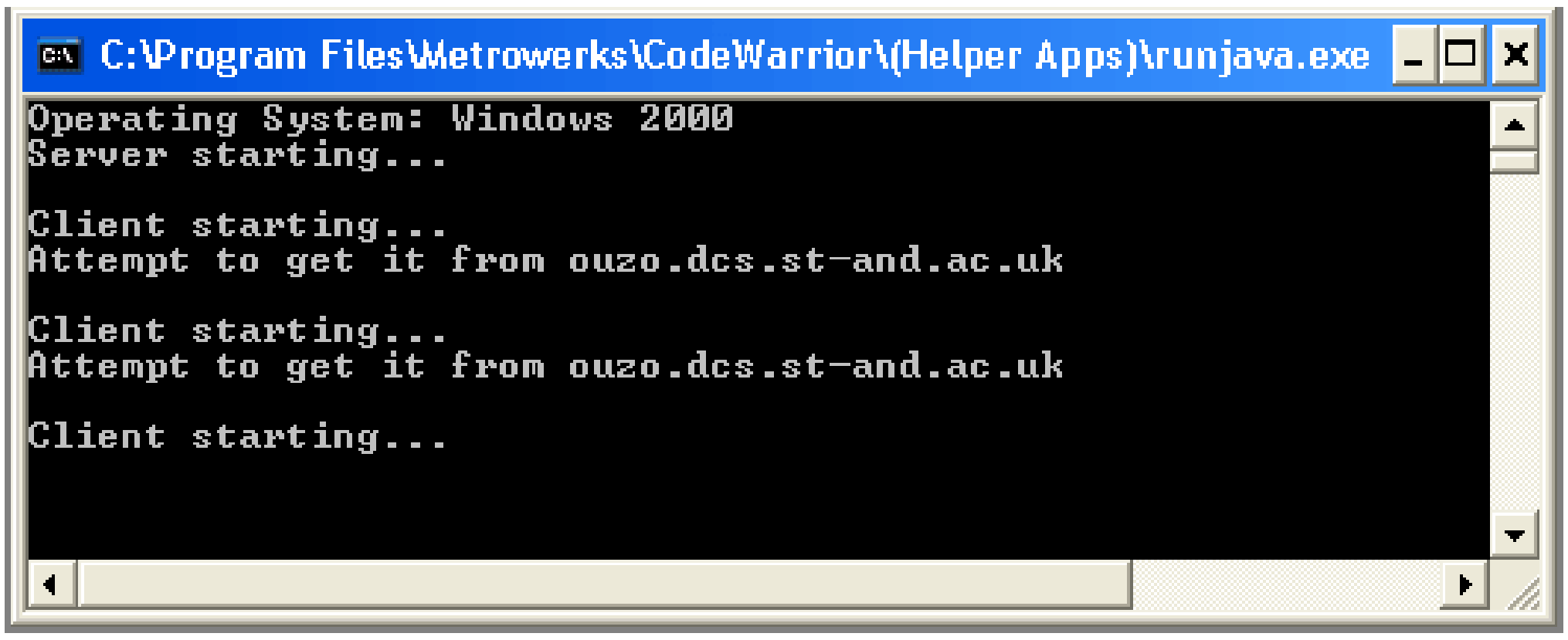

**Figure 34:** Console window messages when performing UC6

## 6.7 Use Case 7 (UC7)

UC7 is a variation of UC6. The aim of UC7 is to demonstrate a different topology of stores in remote hosts. In particular, the process of redirecting requests on stores in three hosts is illustrated. The scenario is as follows:

> "*On node **panda** create a shareable local store. On node **ouzo** create a shareable proxy store that knows about node **panda**. On node **tsipouro** create a shareable proxy store that knows about **ouzo**. Add an XML document on the proxy store on **tsipouro** and retrieve it later on.*"

An abstract diagrammatic description of the use case is shown in Figure 35.

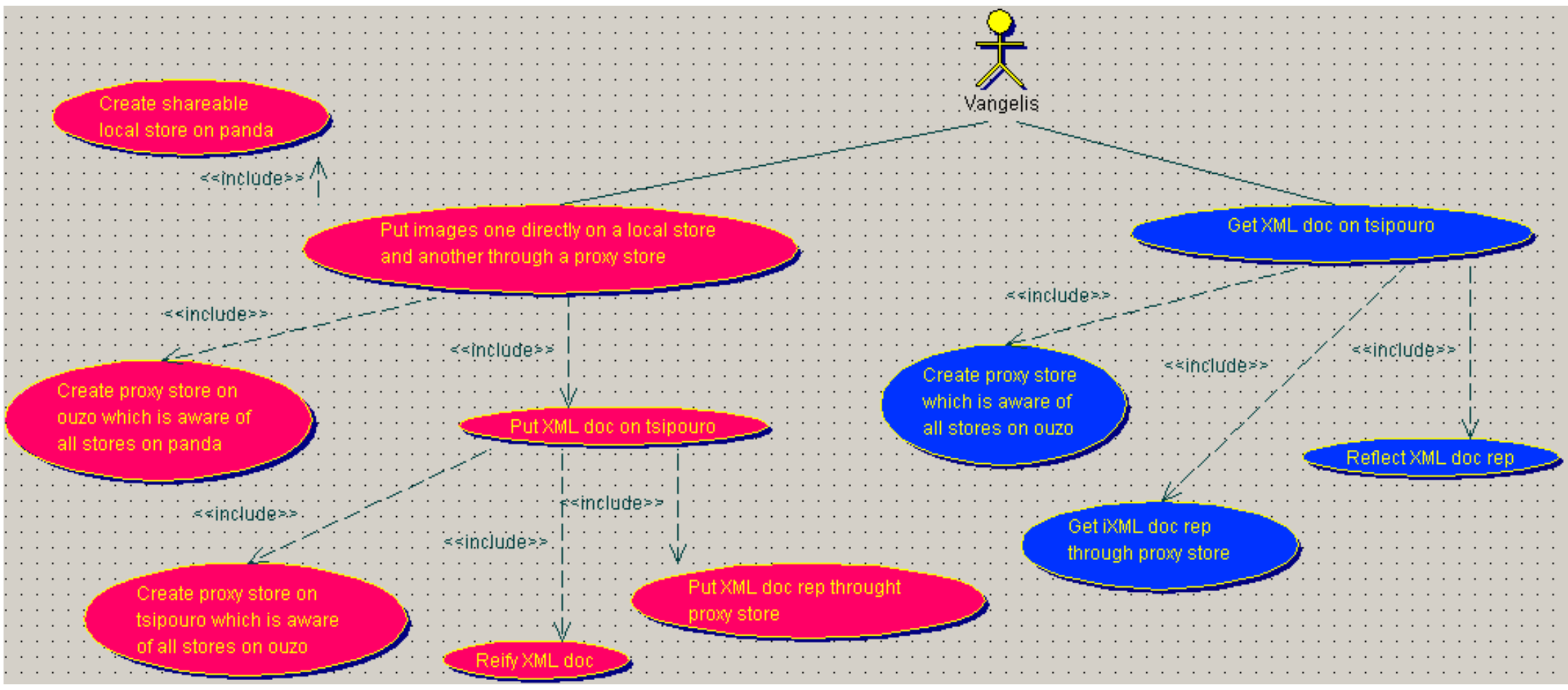

**Figure 35:** Use case diagram for UC7

In order to satisfy the needs of UC5 a caster is required to reflect and reify an XML DOM object. As shown in Figure 17, class *XMLCaster* provides an implementation of such a caster.
The source code for creating a local store on **panda** is shown in Figure 36.

```
// Create a shareable local store
IStore pandaLocalStore = StoreFactory.createLocalStore(true);
```

**Figure 36:** Creating a shareable local store on **panda**

The source code for creating a shareable proxy store on **ouzo** that knows about stores on node **panda** is shown in Figure 37.

```
// Create a shareable local store
IProxyStore ouzoProxyStore = StoreFactory.createProxyStore(true);

// Make that proxy store aware of host panda
// The URL does not specify a particular store, which means that this proxy store will be aware of all the shareable stores in the remote host.
ouzoProxyStore.addTarget( Util.createURL("panda.dcs.st-and.ac.uk") );
```

**Figure 37:** Creating a shareable proxy store on **ouzo**

The source code for achieving the storing part of UC7on **tsipouro** is shown in Figure 38.

```
// Create a DOM document object from a string representation of an XML document
Document xmlDoc = Util.createDOMFromFile("xbaseMembers.xml");

// Create a caster specifically written for XML documents
ICaster xmlDocCaster = new XMLCaster();

// Flatten the XML document into a bitstring
IBitString xmlDocRep = xmlDocCaster.reify( xmlDoc );

// Create a proxy store that knows about node ouzo
IStore aProxyStore = StoreFactory.createProxyStore( Util.createURL("ouzo.dcs.st-and.ac.uk") );

// Put the representation of the XML document in the proxy store
// Behind the scenes:
// since the proxy store is not aware of any local stores, it will redirect the request to the available stores on remote hosts.
// The only available store on a remote host is on ouzo, and this is where the operation is performed.
// Ouzo receives the request. The only store available on ouzo to satisfy the request is a proxy store which redirects the request to panda.
// The operation is performed on panda, which replies to ouzo, and ouzo replies to the original request.
IKey xmlDocKey = aProxyStore.put( xmlDocRep );
```

**Figure 38:** Storing part of UC7 on **tsipouro**

The source code for achieving the retrieval part of UC7 on **tsipouro** is shown in Figure 39

```
// Create a proxy store that knows about node ouzo
IStore aProxyStore = StoreFactory.createProxyStore( Util.createURL("ouzo.dcs.st-and.ac.uk") );

// Retrieve the XML document from the proxy store.
// Behind the scenes:
// since the proxy store is not aware of any local stores, it will redirect the request to the available stores on remote hosts.
// The only available store on a remote host is on ouzo, and this is where the operation is performed.
// Ouzo receives the request. The only store available on ouzo to satisfy the request is a proxy store which redirects the request to panda.
// The operation is performed on panda, which replies to ouzo, and ouzo replies to the original request.
IBitString xmlDocRep = aProxyStore.get( xmlDocKey );

// Create a caster specifically written for XML documents
ICaster xmlDocCaster = new XMLCaster();

// Create an XML document DOM object from the bitstring
Document reflectedXMLDoc = (Document)xmlDocCaster.reflect( xmlDocRep );

// Display the xml document
Util.displayXMLDocument( reflectedXMLDoc );
```

**Figure 39:** Retrieval part on **tsipouro** of UC5

The result of executing the code illustrated above is shown in Figure 22. Figure 40 illustrates messages generated when redirecting requests. The upper windows displays messages on **tsipouro** whereas the lower window displays messages on **ouzo**. The messages on **tsipouro** indicate that request is forwarded to **ouzo**. **Ouzo** receives the message and redirects the request to **panda**. **Panda** sends back to **ouzo** the requested representation, and **ouzo** sends it back to **tsipouro**, where it is manipulated appropriately.

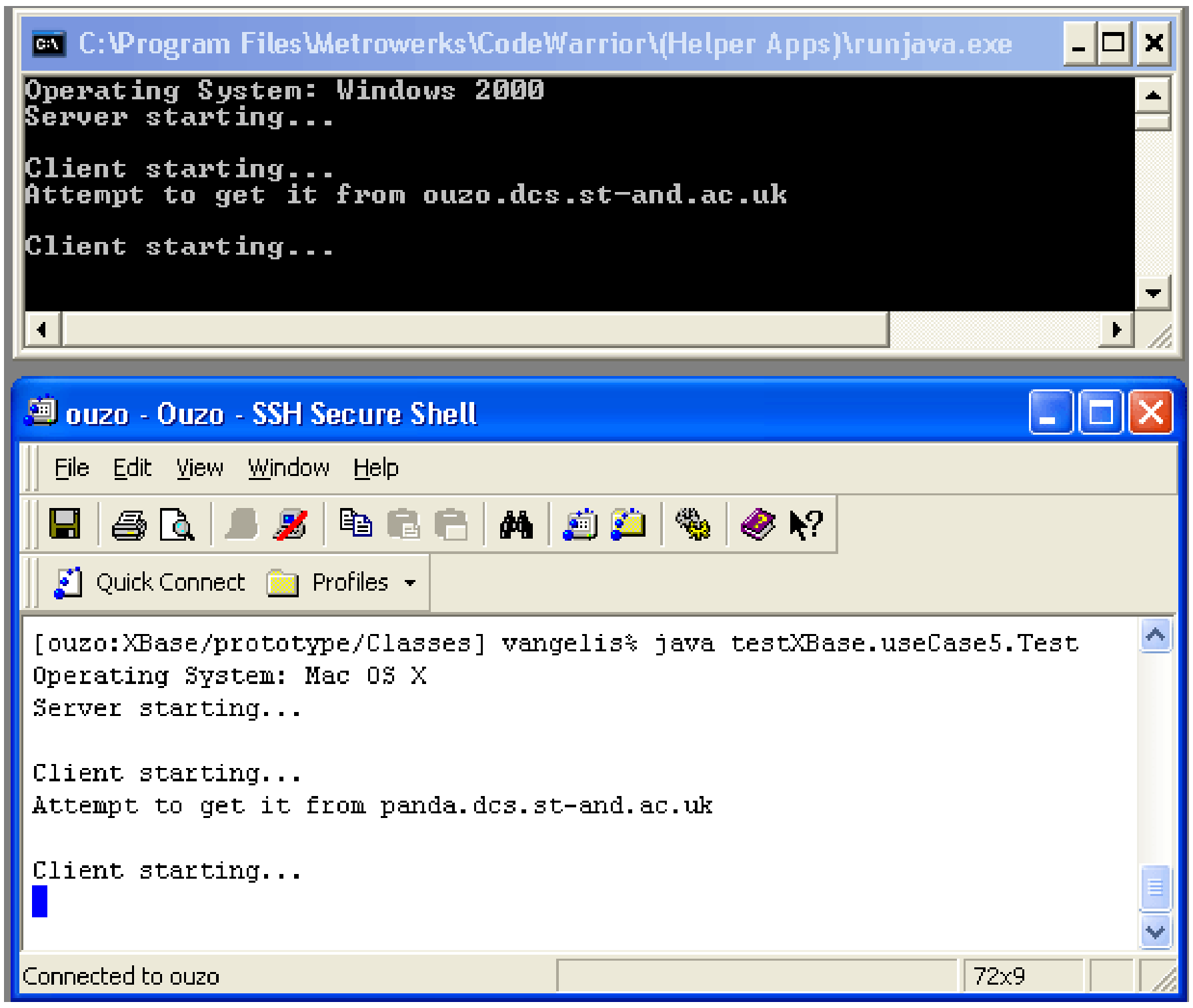

**Figure 40:** Result of UC7

# 7 Failure Semantics

There are several XBase exceptions thrown when constructing XBase components as well as performing several operations. In some cases it is significant for these exceptions to be caught explicitly by the programmer, since it is required to specify what has to be done when a failure occurs. Other exceptions are caught implicitly by the system. The reason for defining such exceptions is to give semantics in the case of particular failures, especially Java system failures. This means that basic Java exceptions are wrapped into XBase defined exceptions. Figure 41shows the set of exceptions used in the XBase system.

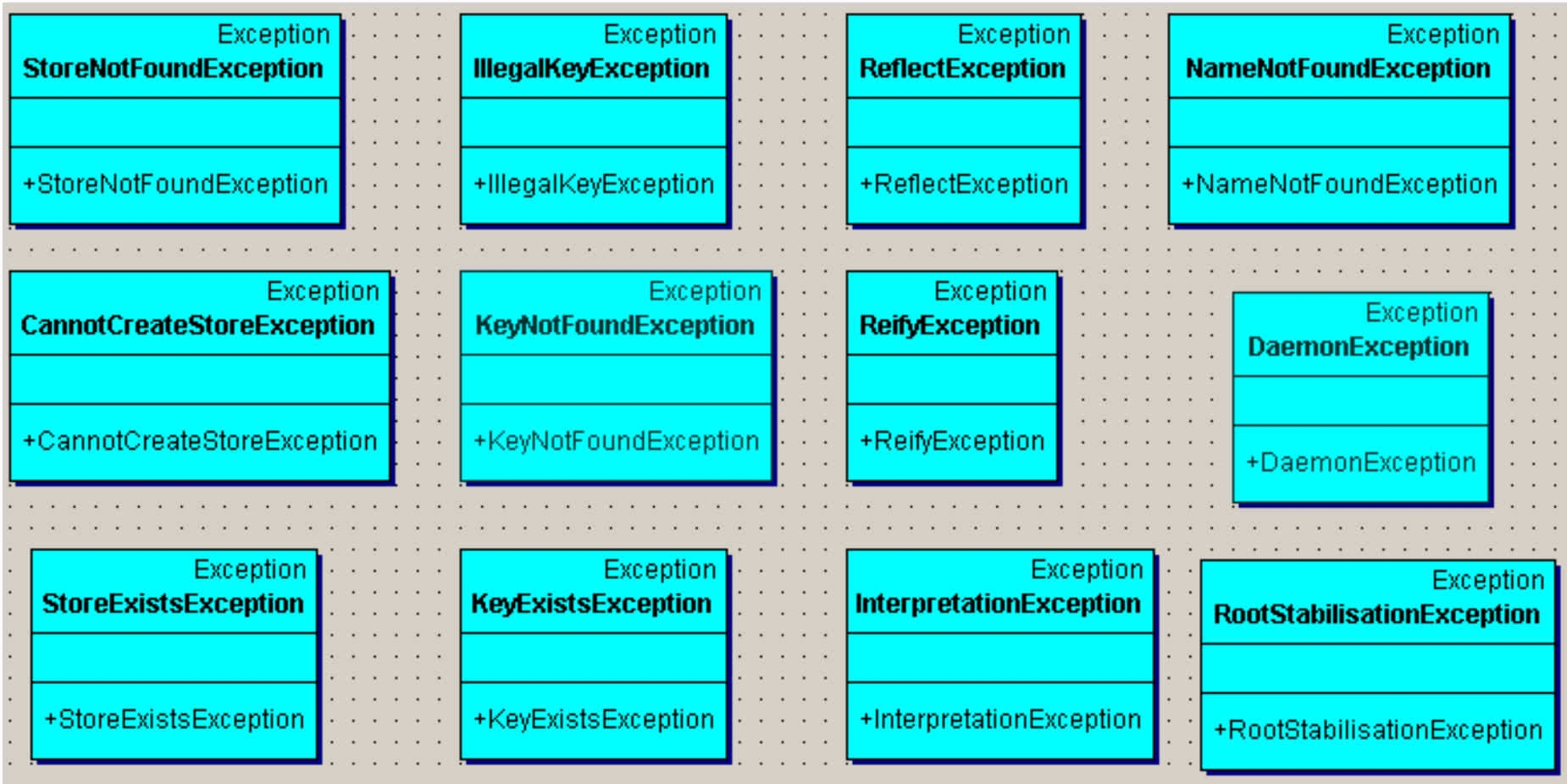

**Figure 41:** Exceptions thrown when composing XBase components

Table 1 contains a short description of the exceptions displayed above. Note that the first four exceptions are caught by the XBase system, whereas the rest are required to be caught by the use case programmer.

| Exception Class | Thrown when the system attempts to… |
|---|---|
| StoreNotFoundException | open an existing store and fails when it does not find one or a network store does not contain any stores at all |
| StoreExistsException | create a new local store that already exists (refers to backing storage) |
| DaemonException | start a daemon and this fails |
| IllegalKeyException | compare two keys that are not of the same format |
| CannotCreateStoreException | create a local store and fails when the underlying backing storage cannot be created |
| KeyNotFoundException | perform a get operation and fails when there is no binding for the given key |
| KeyExistsException | perform a put operation and fails when there is already a binding with the same key |
| ReflectException | translate a representation into a Java object and fails when either an internal error occurs and or the representation is of the wrong format |
| ReifyException | flatten a Java object into representation and fails when either an internal error occurs and or the object cannot be flattened |
| InterpretationException | interpret a representation into another representation and fails when the result is not the desired one |
| NameNotFoundException | find a name in a namer and fails when the request binding cannot be found |
| RootStabilisationException | stabilise a root a fails due to unexpected errors. |

**Table 1:** Exceptions thrown in the XBase system

# 8 Persistence of XML Documents

This section addresses various issues related to the persistence of XML documents using the XBase framework (A description of which is provided elsehere). In particular we should be able to store and retrieve a set of XML documents to/from a persistent repository (e.g. persistent store) using the XBase components described elsewhere. As shown in the following figure, in order to achieve that a set of representations must be created first. These representations can be stored and retrieved to/from a store.

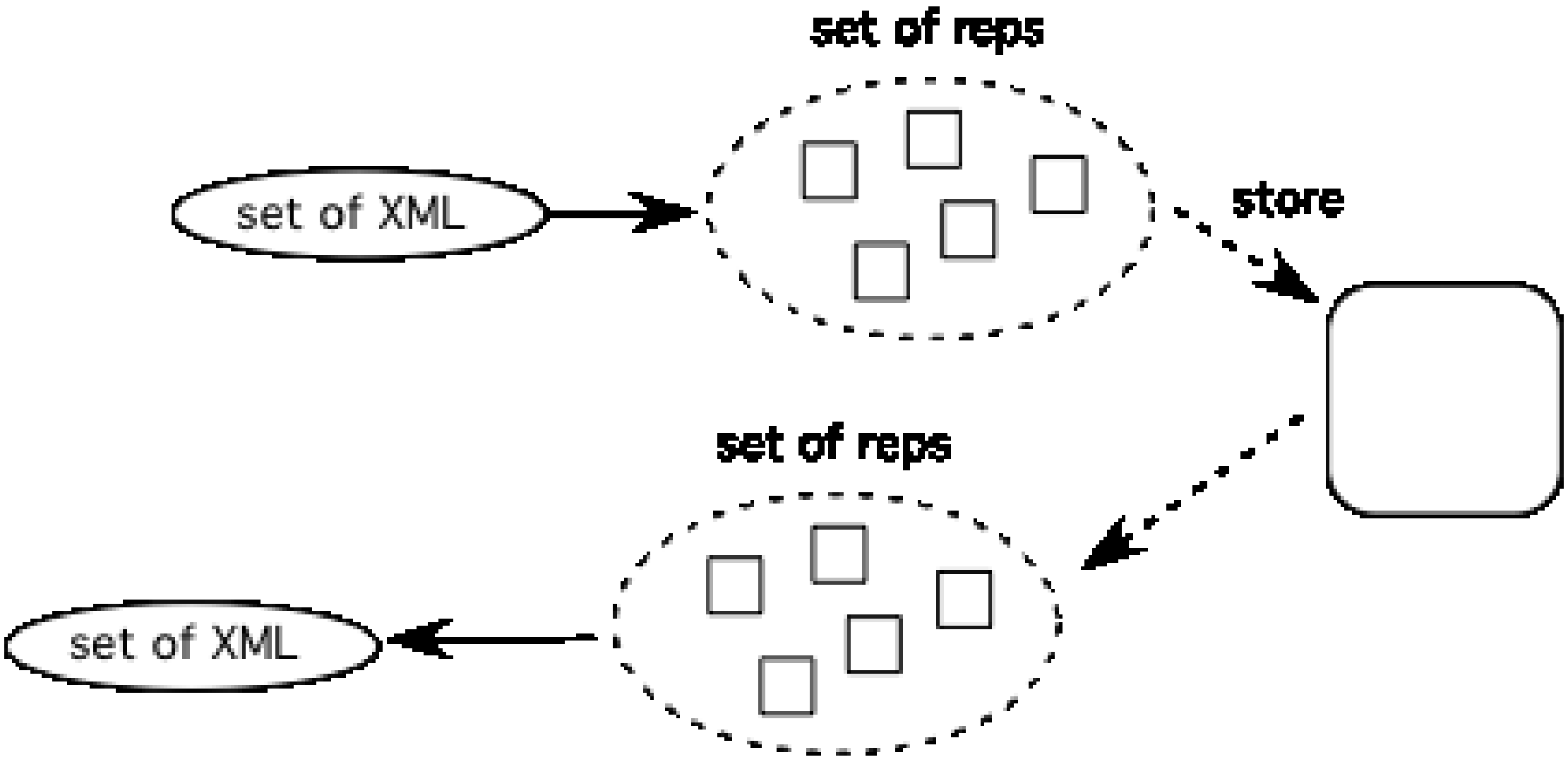

**Figure 42:** Persistence of XML Documents - Use Case

In this section we mainly concentrate on the process of creating representations from entities (XML documents) and vice versa (reification and reflection respectively). This process has been introduced earlier as casting. There are several issues with respect to the form of the representation and the way such a representation is produced. These issues are discussed in the next two sub sections: the first discusses the dimensions in casting XML documents within a distributed XBase system without taking into account the process of update whereas the second describes what changes/additions in the original design should made in order to include updating and versioning as well.

## 8.1 Casting without update

This section discusses the main issues in flattening a set of XML documents into a set of representations and vice versa (reification and reflection respectively). Casting should be performed by taking into account that the resulting representations may be distributed around different stores. However, the main issues here are related to the following dimensions:

- The form of the representation(s) possibly resulting from reification,

- The way that reification and reflection processes are performed. In particular, the way a set of representations is produced when reifying an XML document as well as the way an XML document is constructed when reflecting a set of representations.

The process of casting and effectively the form of the representation depends on several policies that should be taken into account. Some of these policies are user specified whereas some others provide the different options at the implementation level. These policies can be summarised as follows (extended discussion about these policies is included in the next section):

- User Policies
  - **Granularity:** this policy specifies the number of representations resulted from the process of reification as well as the way mapping between these fragments is achieved, given a set of XML documents.

- Implementation Policies
  - **Representation format:** this policy specifies the particular format of a representation of an XML document. An example format is the XML format.
  - **Reference format:** this policy specifies the format of a reference from one representation fragment to another. Possible alternatives are: key, name etc. References here are used in order to denote other representations of XML documents and they are not used in the same way as XML references (e.g. Xlink, Xpointer etc). The policy chosen affects the underlying store communication infrastructure, that is if the reference does not contain any information about the location of the referenced representation, an external mechanism is required to locate it. Similarly, if names are used as references then a policy specifies how names are resolved. Possible alternatives are: user specified namer, fixed position of a namer etc.

The next two sections discuss those policies (user and implementation respectively). This is followed by a definition of a particular set of requirements that will guide us to decide which particular policies are suitable to satisfy those requirements. Finally, a description of a particular implementation of a caster is provided.
For the rest of this document we will assume the following use case: an XML document is reified and this results in a set of representations. These representations are then reflected in order to give us a copy of the original XML document (an example will be given in a later section). In addition, when we refer to an XML document we denote an entity.

### 8.1.1 User Policies

User policies define a set of parameters that are external to the caster. In other words it defines various user specific customisations that make the system compliant to the user needs. These customisations could be applied to any XML document caster and they only change the default behaviour of the caster and not the way it is implemented.

The main issue here is to specify the granularity, that is given an XML document how many fragments are resulted from the process of reifying. In addition, when breaking an XML document into smaller fragments a mapping between these fragments is required. The spectrum of granularity is in between having a single fragment per XML tag and having a fragment for the whole XML document. There is a trade of between efficiency and flexibility as we move from one extreme case to another. Choosing either of the cases is an implementation decision and results in too fine or coarse grained representations respectively.
However, the user may customise the granularity, that is finding the critical point where efficiency and flexibility will be maximised with respect to the particular needs of the application. Customisation requires the user to provide the corresponding set of rules, which can be expressed in the form of XML schemas. The schema defines which elements of the XML document are kept in the same representation. An example of a schema will be given later.

### 8.1.2 Implementation Policies

Implementation policies define a set of different parameters that should be taken into account when designing a caster. These parameters are hard-coded in the caster and consequently hidden from the user. Different set of policies results in different casters.
The first policy is related to the particular format of the representation(s) of an XML document. Any format can be suitable, whether it is intelligible or not. However, this requires from the caster to provide the mechanism in order to interpret it appropriately. An intelligible candidate format for a representation can be a self-contained valid XML document. The usage of XML allows the caster to use already predefined tools to manipulate that format. In addition the usage of a standard format makes it understandable by any user. Using XML as the format for representations means that both the XML documents (entities) and their representations are in XML format. The difference is that a representation may contain references to other representations, which have to be resolved first (an operation performed inside the caster) in order to reflect to the desired XML document entity.
The form that references between representations have is specified by another policy. The process of reifying an XML document may result in more than one representations that reference each other. This is analogous to having hyper-links denoting entities in a hyper-code system.
A reference to any other representation should contain sufficient information which will be used in order to retrieve the referenced representation from anywhere in the distributed environment (that is a local or a remote store). This implies that during casting (reflection in particular), the caster will attempt to resolve the reference.
A possible form of a reference could be a key that will be generated for the referenced representation during the reification process. This form of a reference is suitable only in a system with no update or versioning and in a system with a store that is able to accept a key generated externally (the key is generated during the reification process and if a user puts that representation is a store, this store will not generate a key but will accept the key that is provided by the user).
Another possible form of a reference to a representation of an XML document is a name, which can be accessible either locally or globally. This form of a representation does not require neither a system with no update or versioning nor a system with a store been able to accept a key externally (since the system can generate a name for the referenced representation during the reification process and if a user puts that representation in a store, the key resulted is bound to that name). However, it requires the system to make use of a namer (a data structure that binds names with keys), which brings up the issue of where this namer is located (see policy about namers later on).
However, in either case (key or name) we make the assumption that there is an external mechanism for locating the referenced representation, since this kind of reference contains no information about the location. This means that the underlying store communication infrastructure is sufficient to be able to discover representations from given keys. The mechanism for making stores knowing about each other can be plugged in the XBase system (e.g. a Pastry peer-to-peer).
An alternative to that is to have a form of reference that contains information about the location of the referenced representation, which means that the system does not depend on the store communication scheme. However, this makes the representation location dependant and the system less flexible especially if a system supports update and versioning.
In any case, the fact that the representation that corresponds to a reference may contain more references to other representations, which require resolution as well, makes the reflection process a recursive operation. In addition, the operation of resolving references may lead to cycles, which are detected as casting goes along by keeping track of each representation (more details later on).
Another point is that a reference has different semantics from an XML reference (e.g. XLink, XPointer) and the difference is that the former corresponds to a representation that has to be resolved first, whereas the latter is part of the original XML document (entity).
In the case of making use of a name as a reference, we have to specify how these names are resolved. Using the XBase framework this problem can be solved by using namers. The policy here is related to locating namers that contain the name/key bindings. This is required by a caster when attempting to resolve the name. Choosing an alternative for that policy is a trade of between efficiency and flexibility.
One possibility is to make use of a user specified namer. Although this possibly allows fast resolving (especially in the case where the namer is used locally), however it does not automate the process of resolving names, since it requires from the user to specify the location where the namer resides.
Another alternative is to make use of a namer or a set of namers, which are located in a fixed place (e.g. making use of a root namer). In this case, the system attempts to automatically find the appropriate root namer and resolve the name, however making the process less flexible, due to the fact that the namer is kept is a fixed location.

### 8.1.3 Requirements for a distributed XBase system with no update or versioning

Several general requirements for designing an XBase system have been specified in an earlier document. These requirements were: simplicity, flexibility, efficiency and portability. XML document casting is guided by these dimensions. This section describes a more concrete set of requirements that extend the original requirements and guide us choose a particular implementation of a caster. These requirements are:

- The system must allow distributed information. This means that fragments or representations may be scattered around in different places and the caster should be able to find that information as efficient as possible

- The representation chosen for XML documents should be in an intelligible standard form. The user should be able to customise the granularity as much as possible (this has been described in a previous section) in order to make it compliant to his needs.
- References to other representations of XML documents should contain enough information in order to retrieve the referenced entity from anywhere in the system (e.g. a store) in the most flexible way. The restrictions imposed by a particular implementation should be as much as possible minimal.

### 8.1.4 Meeting the requirements (Choosing particular policies)

The requirements specified in the previous section allow us to define a particular set of policies in order to produce a particular XML document flattening technique (and effectively to provide a particular implementation of a caster. A set that satisfies the criteria specified above is:

- The representation is in XML format. Reifying an XML document results in a set of fragments that may reference each other. The user specifies the rules (in the form of XML schema) that will instruct the caster about the granularity when it is attempted to normalise an XML document. The rule is that a separate fragment will be created for each reference in the schema (See XML schema *ref* attribute.
- The reference is a name that has to be resolved during casting. Resolution is a two stage process: first the name is to be presented to a user specified namer in order to get the key which is then presented to the underlying storage infrastructure in order to retrieve the requested representation.

### 8.1.5 A Proposed XML Document Casting Technique

In this section we describe a particular implementation of a caster that satisfy the policies described above. This involves the definition of the interfaces as well as a description of the algorithms.
Interface Caster, defined elsewhere, should be modified accordingly since reifying an XML document may result in more than one representation. Similarly, the reflect operation should be able to accept the result of the reification process, that is a set of representation fragments. Therefore, the original definition of interface caster described elsewhere should be modified to the following (t is the XML document entity that contains both the XML document and an XML schema):

**reify**: *t* -> set[*BitString*] *or error*
**reflect**: set[*BitString*]-> *t or error*

#### 8.1.5.1 Description of the Algorithms

Below we describe the algorithms for both operations. The policies that guide the particular implementation are:

- the representation is in XML format,
- the reference is a name and the namer used to resolve it is in the form of a root namer in a fixed location.

The algorithm for the reify operation is shown in Figure 43:

- Extract the rules of granularity, (variable *schema*), included in the given entity.
- Extract the xml document to be reified (variable *xml*), included in the given entity.
- If xml==null then throw an error
- If schema==null then , If policy is to throw an error, then throw it else generate a default schema.
- Validate the document against the schema. If this fails then throw an error
- Start parsing the schema, recursively scan all the elements
- For each element tag that references another element tag record the XPath expression and record it in a vector (variable *paths*)
- Start parsing the xml document, recursively scan all the element.
- For each element tag{
    - If the current element's XPath expression exists in *paths* then {
        - Replace that element with a reference tag that has the general form <XBaseRef ref="*(system generated ID)*"> (SYS–ID1).
        - Create a new document by wrapping that element within a new tag that has the general form <XbaseName ref="*(the ID generated before)*">.
    - }
- }
- Wrap the outer most tag within a new tag that has the general form <XbaseName ref="*(system generated ID SYS-ID2)*" schemaRef="(SYS-ID2)-schema">
- Wrap the schema fragment within a new tag that has the general form <XbaseName ref="*(SYS_ID2)-schema*">
- return the resulting bitstrings

**Figure 43:** An algorithm for the *reify* operation

The way that the Ids (both SYS-ID1 and SYS-ID2) are generated by the system is implementation specific. A general pattern for both kinds of Ids could be:

- SYS-ID1: (Node IP address)+(Element full path and name)+(counter)
- SYS-ID2: (Node IP address)+(Element full path and name)+(outer most tag keyword)+(counter)

For example:

- SYS-ID1: 192.0.0.1-XBaseMembers/researchFellows/person-1
- SYS-ID2: 192.0.0.1-XBaseMembersOuterMostTag-1

Note that the above algorithm consists of two parts. The first part scans a schema definition in order to extract XPaths. These XPaths are used in the second part of the algorithm in order to create the different fragments. Consequently an alternative implementation of the reify operation could be to ask from the user to specify XPaths rather than the schema, which sometimes could be hard or could require a sophisticated User Interface.

An algorithm for the reflect operation is shown in Figure 44:

- For each given fragment {

- If it is well formed then extract the name and the document that is encapsulated, and record that tuple in *tuples*.
- }
- if there is at least one invalid fragment then throw an error
- Extract the fragment that contains the outer most tag (variable *rootDoc*)
- If *rootDoc*==null then throw an error
- Extract the fragment that contains the schema (variable *rootSchemaDoc*)
- Starting from *rootDoc*, recursively scan all the elements.
- For each reference in the fragment (XBaseRef element) {
    - Extract the name from the reference
    - If there is a fragment among the given that is identified by a name that is equal with the name included in the reference {
        - Replace the reference tag with the element encapsulated in the fragment
    - } else throw an error
    - Recursively scan the children of the element for references and perform the same substitution task.
- }
- Create an XML entity that couples the resulted XML document and the schema (both *rootDoc* and *rootSchemaDoc*)
- Return that entity

**Figure 44:** An algorithm for the *reflect* operation

The algorithm described above requires the user to provide all the necessary fragments as parameters. This means the system only looks among the given for a requested fragment. An alternative approach is to make use of a root namer in order to find a request fragment. The highlighted lines in Figure 44, will be replaced by the piece of code shown in Figure 45:

- If there is a fragment among the given that is identified by a name that is equal with the name included in the reference then keep it for later use (variable *fragment*)
- else if the root namer contains a binding with a name that is equal to the name included in the reference element then {
    - Find the corresponding key by resolving the name in the root namer
    - Present that key to the root store.
    - If there is a fragment as a result then keep it for later use (variable *fragment*)
    - else throw an error.
    - Replace the reference tag with the element encapsulated in *fragment*
- } else throw an error

**Figure 45:** Resolving a name reference in the *reflect* operation

Using the second algorithm for resolving a name provides more flexibility in the process of storing and retrieving XML documents. The only thing that the user has to remember when reifying an XML document is the name of the fragment that encapsulates the outer most element. Use case 2 demonstrates the way that can be done.

#### 8.1.5.2 An Example

In order to describe how the proposed caster implementation works, we are going to reify the XML document shown in Figure 46, which describes some various information about the members of the XBase project (Note that this is not a description of a use case. It only shows what is the result of reification for a particular example):

```
<XBaseMembers>
    <researchFellows>
        <person>
            <name> Evangelos Zirintsis</name>
            <address>
                <town>St Andrews</town>
            </address>
            <age>29</age>
        </person>
    </researchFellows>
    <teachingStaff>
        <person>
            <name> Graham Kirby</name>
            <address>
                <town>Kingsbarns</town>
            </address>
            <age>36</age>
        </person>
        <person>
            <name> Ron Morrison</name>
            <address>
                <town>St Andrews</town>
            </address>
        </person>
        <person>
            <name> Al Dearle</name>
            <address>
                <town>St Andrews</town>
            </address>
        </person>
    </teachingStaff>
</XBaseMembers>
```

**Figure 46:** An example XML document

A schema that the example XML document complies is shown in Figure 47:

```
<xsd:schema xmlns:xsd=”http://www.w3.org/2000/10/XMLSchema”>
    <xsd:element name = “XBaseMembers” />
</xsd:schema>
```

**Figure 47:** A schema definition that corresponds to the XML document displayed in **Figure 46**

The above schema can be used in order to specify the granularity. In this particular case no separate fragments will be created. Another example of a schema is shown in Figure 48:

```
<xsd:schema xmlns:xsd="http://www.w3.org/2001/XMLSchema">
  <xsd:element name="XBaseMembers">
    <xsd:complexType>
      <xsd:sequence>
        <xsd:element name="researchFellows">
          <xsd:complexType>
            <xsd:sequence>
              <xsd:element ref="person" minOccurs = "0" maxOccurs = "unbounded" />
            <xsd:sequence>
          </xsd:complexType>
        </xsd:element>
        <xsd:element name=" teachingStaff ">
          <xsd:complexType>
            <xsd:sequence>
              <xsd:element ref="person" minOccurs = "0" maxOccurs = "unbounded" />
            <xsd:sequence>
          </xsd:complexType>
        </xsd:element>
      </xsd:sequence>
    </xsd:complexType>
  </xsd:element>

  <xsd:element name="person">
    <xsd:complexType>
      <xsd:sequence>
        <xsd:element name = "name" />
          <xsd:element ref="address" />
          <xsd:element name="age" minOccurs = "0" />
        </xsd:sequence>
    </xsd:complexType>
  </xsd:element>

  <xsd:element name="address" />

</xsd:schema>
```

**Figure 48:** An alternative schema definition

The definition of that schema effectively groups together the details of each person separately. In addition, information about address is grouped together in a separate fragment.
The reification process included in a caster takes into account the schema and transforms the original XML document into the following representation;

```
<XBaseName   ref="192.0.0.1-XbaseMembersOuterMostTag-1"
             schemaRef="192.0.0.1-XbaseMembers-OuterMostTag-schema-1">
   <XBaseMembers>
       <researchFellows>
           <XbaseRef ref="192.0.0.1-XBaseMembers/researchFellows/person-1" />
       </researchFellows>
       <teachingStaff>
           <XbaseRef ref="192.0.0.1-XBaseMembers/teachingStaff/person-1">
           <XbaseRef ref="192.0.0.1-XBaseMembers/ teachingStaff /person-2">
           <XbaseRef ref="192.0.0.1-XBaseMembers/ teachingStaff /person-3">
```

```
        </teachingStaff>
    </XBaseMembers>
</ XBaseName>
```

**Figure 49:** The transformed version of the original XML document

In this representation the *person* elements have been replaced by a tag that specifies a name, which acts as a reference. The fragments for each of the *person* elements are kept separately. In the XML shown in Figure 50 we show the fragment for the first person element (The fragments of other person elements have similar format). This fragment is wrapped in a tag which contains information about the name given by the system.

```
<XBaseName ref=”192.0.0.1-XBaseMembers/teachingStaff/person-1”>
    <person>
        <name> Evangelos Zirintsis</name>
        <XBaseRef ref=”192.0.0.1-XBaseMembers/teachingStaff/person/address-1” />
        <age>29</age>
    </person>
</ XBaseName>
```

**Figure 50:** A fragment containing the details of a person

The fragment above contains a reference to an the specified address element which is shown in Figure 51:

```
<XBaseName ref=”192.0.0.1-XBaseMembers/teachingStaff/person/address-1”>
    <address>
        <town> St Andrews</town>
    </address>
</ XBaseName>
```

**Figure 51:** A fragment containing address details

## 8.2 Flattening with update and versioning

This section describes the extra dimensions of a system that supports updating and versioning.
The first policy is related to XML document updating. A sub-policy is to define the update model, that is categorising the different kinds of update that can be performed on an XML document and decide which ones will be supported by the current XBase implementation. The first main category includes updates to an XML document that results in an XML document that conforms to the original schema. This involves modification of tag values, addition and deletion of nodes. The current system fully supports this kind of update.
The second category includes any updates that result in an XML document that does not conform to the schema included in the XML entity (entity *t* as defined earlier). In addition to modification of tag values, addition and deletion this involves modification of nodes as well. This kind of update is supported in the current system as long as the user provides the new schema that the updated XML document conforms to.
Another sub-policy concerning XML document updating is related to the XML difference discovering mechanism. This is to specify the way that the system discovers the differences between different versions of XML documents. A possible decision is to have the caster in charge of discovering which XML documents have been updated. This is done by comparing the original set of fragments that consist the representation of an XML entity with the fragments resulting after making any changes. This results in possible differences which denote updated XML fragments. Another alternative is to present the fragments resulting after making changes to the store. The store produces a key which is then compared to the existing keys. If the key already exists in the store this means that the fragment has not been changed. This alternative takes advantage of the fact that the keys generated are content-based.
Another policy is related to defining the identity of individual nodes in an XML document. In the update model defining identities is necessary in order for the system to decide which fragment is the updated version of which original fragment. The policy here is to decide how to provide identity. One alternative is to insert an ID attribute in each element, which means that the original XML document has to be modified. Another alternative is to keep an external mapping between IDs and nodes (see later section).
Another policy is related to supporting historical versions, In particular this refers to the information stored. It has two branches. The first is what information should be stored and the second where to store it. Possible alternatives for the former are: a timestamp automatically generated by the system, a version ID, provided by the user.
A possible alternative for the latter is to store it inside a namer, which means that the user has to provide version information together with the name.

### 8.2.1 Extra components/modifications required

Entity *t* defined earlier has to be modified in order to include information about node identity. In particular, apart from the XML document and the corresponding schema, entity t contains the mapping between ids and nodes included in the XML document. Each node has a unique name which is used as id for as long as this node is "alive".
In order to support update and historical versions, the namer should support resolving a name together with version information. The interface for namers in that case should be the following (Note that VersionTuple couples together a name and version information):

**lookup**: *VersionTuple* -> *Key* or *error*
**bind**: *VersionTuple*, *Key*
**unbind**: *VersionTuple*, *Key*

Use Case 3 describes how to use the new components in order to perform an update.

## 8.3 Use Cases

In this section we provide particular examples of composing the XBase components described in another document, in order to provide persistence of XML documents. The use cases presented here demonstrate:

- Usage of a caster for the persistence of a particular XML document – The user has to know about all the keys (Use Case 1)
- Usage of a root store and root namer for the persistence of a particular XML document – The user has to know only about the name of the fragment of the top level tag. A user interface is also used to demonstrate the way the user can specify granularity (Use Case 2).

- Usage of extra/modified components required to support update and historical versions (Use Case 3).

The sections that follow contain descriptions of various scenarios and their corresponding Java source code that shows the way that XBase components may be composed to satisfy the scenario.

Note that the code demonstrated is split into two levels. The first level is the end user level, that is the high level API used to satisfy the scenario. The second level is the XBase user level, that demonstrates the way XBase components may be composed.

### 8.3.1 Use Case 1 (UC1)

The aim of UC1 is to demonstrate the usage of a caster in order to put/retrieve an XML document to/from a store respectively. The root store is a proxy store that knows about a local store. The scenario is as follows:

> "*On node **tsipouro** add an XML document in a local store which is accessible through the root store (root store is proxy store). Using the resulting keys retrieve the fragments and present them to the caster in order to reconstruct the XML document.*"

The end user code of UC1 is shown in Figure 52.

```
// Create a DOM document object from a string representation of an XML document
Document xmlDoc = Util.createDOMFromFile("examplesDir/xbaseMembers.xml");

// Create a DOM document object from a string representation of an XML schema
Document xmlSchema = Util.createDOMFromFile("examplesDir/xbaseMembers.xsd");

// Create an XML entity instance that couples the document and the schema.
// IDs are initially empty.
IXMLEntity xmlEntity = new XMLEntity(xmlDoc, xmlSchema);

// Put the entity in the store
IKey[] keys = putXMLEntity( xmlEntity );

// Using the keys retrieve the XML entity from the store.
IXMLEntity retrievedXMLEntity = getXMLEntity( keys );

// Display the xml document
Util.displayXMLDocument( retrievedXMLEntity.getXMLDocument );
```

**Figure 52:** The end user code of UC1

The source for achieving the storing part of UC1 (method *putXMLEntity*) is shown in Figure 53.

```
// Set the type of the root store to be a proxy store
XBase.setRootStoreProperty("proxy");

// Retrieve an instance of the root store
IStore rootStore = XBase.getRootStore();

// Create a shareable local store
IStore myLocalStore = StoreFactory.createLocalStore(true);
```

```
// Make the local store reachable from the root store (root store is a proxy store)
rootStore.addTarget(myLocalStore);

// Create a caster specifically written for XML documents
IXMLCaster xmlDocCaster = new XMLCaster();

// Flatten the XML document into a set of bitstrings
IBitString[] xmlDocRep = xmlDocCaster.reify(xmlEntity);

// Put the representation of the XML document in the root store and record the keys.
IKey[] theKeys = new Key[xmlDocRep.length];
for (int i=0; i<theKeys.length; i++) theKeys[i] = rootStore.put( xmlDocRep[i] );

return keys;
```

**Figure 53:** The storing part of UC1

The source for achieving the retrieval part of UC1 (method *retrieveXMLEntity*) is shown in Figure 54.

```
// Retrieve an instance of the root store
IStore rootStore = XBase.getRootStore();

// Retrieve the fragments from the store
IBitString[] xmlDocRep = new BitString[ theKeys.length ];
for (int i=0; i<theKeys.length; i++) xmlDocRep[i] = rootStore.get( theKeys[i] );

// Create a caster specifically written for XML documents
IXMLCaster xmlDocCaster = new XMLCaster();

// Cast the representation into an entity
IXMLTuple xmlDocAgain = xmlDocCaster.reflect(xmlDocRep);

return xmlDocAgain;
```

**Figure 54:** The retrieval part of UC1

The result of executing the code shown in Figure 52 is shown in Figure 55.

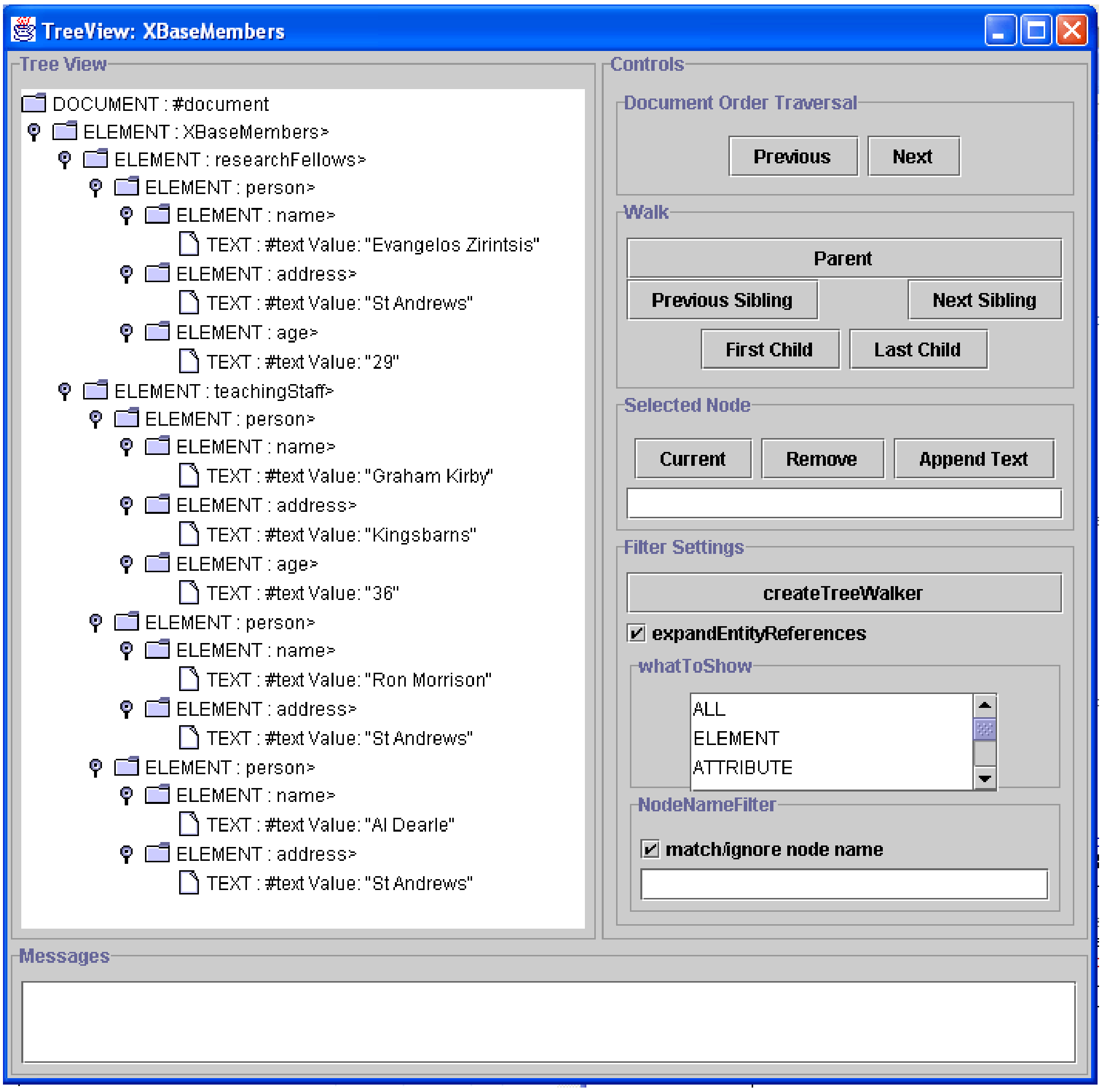

**Figure 55:** The result of Use Case 1

### 8.3.2 Use Case 2 (UC2)

The aim of UC2 is to demonstrate the usage of a root namer in order to put/retrieve an XML document to/from a store respectively as well as demonstrate the way the user can specify the granularity rules. The root store is a proxy store that knows about a local store. The root namer is retrieved and stored from/to the proxy store. The scenario is as follows:

> "*On node **tsipouro** add an XML document in a local store which is accessible through the root store. Retrieve the XML document from the proxy store using a pre-recorded name.*"

The end user code of UC2 is shown in Figure 56.

```
// Create a DOM document object from a string representation of an XML document
Document xmlDoc = Util.createDOMFromFile("examplesDir/xbaseMembers.xml");

// Retrieve the XML schema
```

```
Document xmlSchema = getGranularity( xmlDoc )

// Create an XML entity instance that couples the document and the schema.
// IDs are initially empty.
IXMLEntity xmlEntity = new XMLEntity(xmlDoc, xmlSchema);

// Put the entity in the store
IName xmlDocName = putXMLEntity( xmlEntity );

// Using the keys retrieve the XML entity from the store.
IXMLEntity retrievedXMLEntity = getXMLEntity(xmlDocName);

// Display the xml document
Util.displayXMLDocument( retrievedXMLEntity.getXMLDocument );
```

**Figure 56:** The end user code of UC2

The source for having the user specify the granularity rules (method *getGranularity*) is shown in Figure 57. Note that any tool can be used in order to generate the schema. In this particular use case we are using the XSDInference by IBM [REFERENCE].

```
// Generate the most general schema given the XML document
Document generalSchema = generateSchema( xmlDoc );

// Now open a tree walker view with that schema.
TreeWalkerView schemaTree = new TreeWalkerView( generalSchema );

// Wait for input from the user. When it is done then extract the schema and return it.
Document granularitySchema = schemaTree.getXMLDocument();
```

**Figure 57:** Specifying granularity

The source for achieving the storing part of UC2 (method *putXMLEntity*) is shown in Figure 58.

```
// Set the type of the root store to be a proxy store
XBase.setRootStoreProperty("proxy");

// Retrieve an instance of the root store and the root namer
IProxyStore rootStore = XBase.getRootStore();
INamer rootNamer = XBase.getRootNamer();

// Create a shareable local store
IStore myLocalStore = StoreFactory.createLocalStore(true);

// Make the local store reachable from the root store (root store is a proxy store)
rootStore.addTarget(myLocalStore);

// Create a caster specifically written for XML documents
IXMLCaster xmlDocCaster = new XMLCaster();

// Flatten the XML document into a set of bitstrings
IBitString[] xmlDocRep = xmlDocCaster.reify( xmlEntity );

// Put the representation of the XML document in the root store and record the outer most tag name.
IKey aKey;
```

```
IName aName;
IName recordedName;
for (int i=0; i<xmlDocRep.length; i++) {

        // Put the particular fragment in the store
        aKey = rootStore.put( xmlDocRep[i] );

        // Read the fragment and extract the name.
        aName = XMLGlobals.extractNameFromRep( xmlDocRep[i] );

        // Bind the name with the key
        if (aName != null) rootNamer.bind(aName, aKey);

        // If the fragment represents the outer most tag then record its name.
        if (XMLGlobals.rootTag(xmlDocRep[i]) ) recordedName = aName;
}
return recordedName;
```

**Figure 58:** The storing part of UC2

The source for achieving the retrieval part of UC2 (method *getXMLEntity*) is shown in Figure 59.

```
// Retrieve an instance of the root store and the root namer
IStore rootStore = XBase.getRootStore();
INamer rootNamer = XBase.getRootNamer();

// Resolve the recorded name
IKey xmlKey =rootNamer.lookup( recordedName );

// Retrieve the fragment containing the outer most tag from the store
IBitString xmlFragment = rootStore.get( xmlKey[0] );

// Having the fragment find whats the name of the fragment that represents the schema.
IName schema Name = XMLGlobals.extractSchemaNameFromRep( xmlFragment );

// Resolve that name
IKey[] schemaKey = rootNamer.lookup( schemaName );

// Present the key to store in order to get the schema fragment
IBitString schemaFragment = rootStore.get( schemaKey[0] );

// Create a caster specifically written for XML documents
IXMLCaster xmlDocCaster = new XMLCaster();

// Cast the representation into an entity.
// The caster is only presented with one fragment which means that any other fragments required
// will be retrieved through the root store during the casting process
IBitString[] fragments = { xmlDocRep, schemaFragment};
IXMLTuple xmlDocAgain = xmlDocCaster.reflect( fragments );

return xmlDocAgain;
```

**Figure 59:** The retrieval part of UC2

The root namer is retrieved using function *getRootNamer*. The source in Figure 60 demonstrates the steps followed for retrieving the root namer.

```
// Check whether we have already created a value for the root namer
```

```
if (XBase.rootNamer == null) { // It seems not

        // Retrieve the representation of the root namer.
        IBitString rootNamerRep = XBase.getRootNamerRep();

        // Create a caster specifically written for namers
        ICaster namerCaster = new NamerCaster();

        // Create a namer entity from the representation and record it.
        XBase.rootNamer = namerCaster.reflect( rootNamerRep );
}
return XBase.rootNamer;
```

**Figure 60:** Retrieving the root namer

The following figures illustrate the user interface for specifying the granularity. Figure 61 illustrates the most general schema generated by a tool for the given XML document.

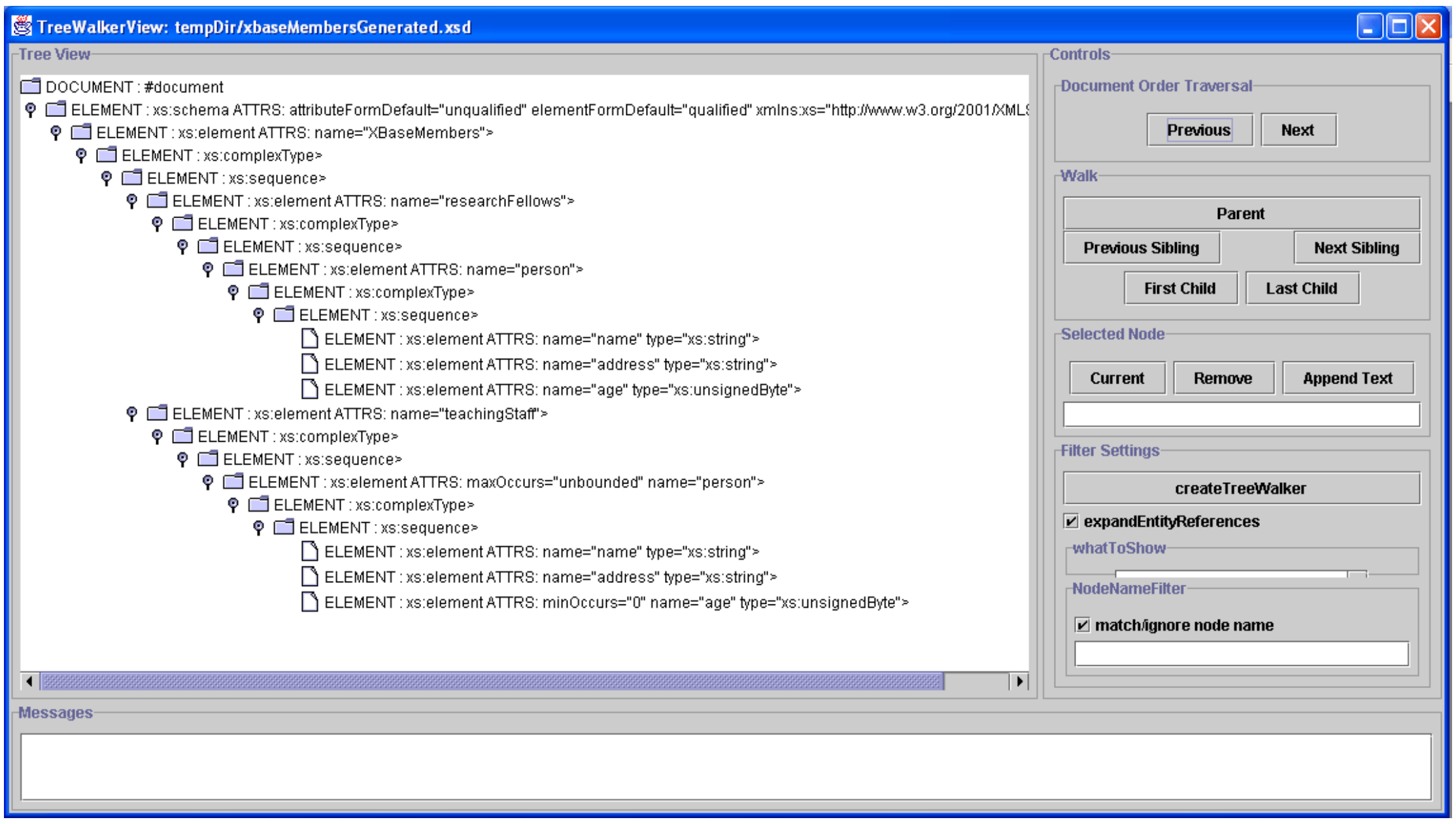

**Figure 61:** The most general schema generated in UC2

In Figure 62, the user has separated the nodes in the schema that will define granularity. For example, notice that line 9 was replaced by a reference to a high-level node which is specified in line 14.

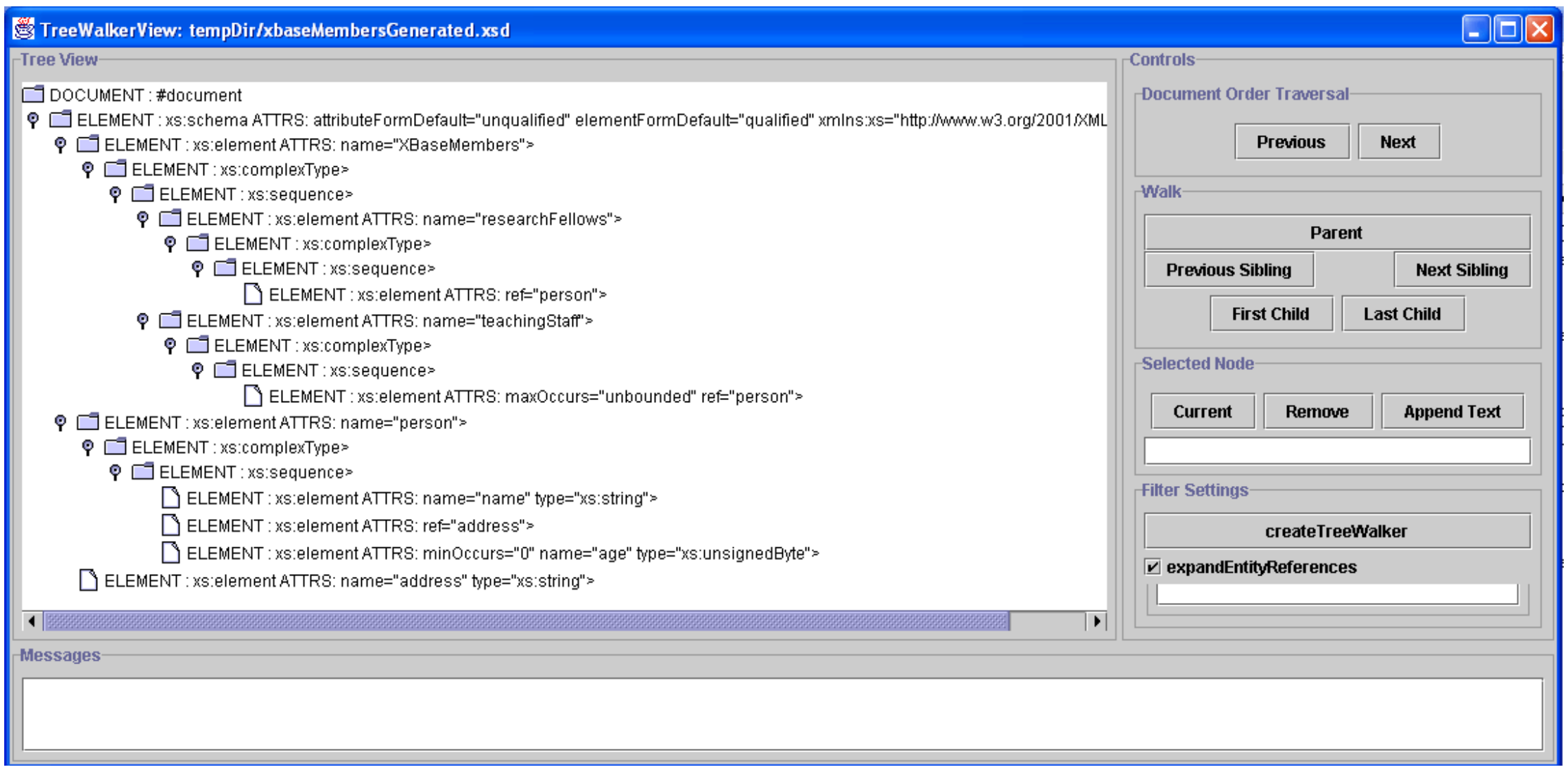

**Figure 62:** Specifying granularity

### 8.3.3 Use Case 3 (UC3)

The aim of UC3 is to demonstrate the usage of a root namer that supports different versions of XML documents/fragments, in order to put/retrieve an XML document to/from a store respectively. The root store is a proxy store that knows about a local store. The root namer is retrieved and stored from/to the proxy store. The scenario is as follows:

> "*On node **tsipouro** retrieve an XML document from the root store. Apply an arbitrary numbers of modifications to that XML document and put it back to the store. Retrieve that XML document using a pre-recorded name and display it.*"

The end user code of UC3 is shown in Figure 63. (It is assumed that the user has already put an XML document in the store and has recorded the name of the outer most tag (name *myXMLDocument*) ).

```
// Retrieve the XML document using the given name as shown in the previous use case.
IXMLTuple xmlEntity = getXMLDocument( new Name( myXMLDocument ) );

// Apply an arbitrary number of changes to that document.
changeXMLDocument( xmlEntity );

// Now perform an update by adding in the store only the modified or new fragments.
updateXMLDocument( xmlEntity );

// Display the updated document
Util.displayXMLDocument( (Document)xmlEntity.getXMLDocument() );
```

**Figure 63:** The end use code of UC3

The source code for updating the entity (method *updateXMLDocument*) is shown in Figure 64.

```
// Create a caster for XML documents.
IXMLTuple xmlEntity = getXMLDocument(xmlDocName);

// Reify the given entity.
IBitString[] fragments = xmlCaster.reify( xmlEntity );

// Retrieve an instance of the root store and the root namer
IProxyStore rootStore = (IProxyStore)XBase.getRootStore();
INamer rootNamer = XBase.getRootNamer();

// Attempt to put the fragments in the store. If the store complaints about a key already present then do
// nothing
IKey aKey;
IName aName;
for (int i=0; i<fragments.length; i++) {
  try {
        // Put the particular fragment in the store
        aKey = rootStore.put(fragments [i] );

        // Read the fragment and extract the name.
        aName = XMLGlobals.extractNameFromRep(fragments [i] );

        // Bind the name with the key
        if (aName != null) rootNamer.bind(aName, aKey);
  } catch (KeyExistsException kee) { }
}

XBase.stabiliseRoot();
```

**Figure 64:** The source code for updating fragments

The particular implementation relies of the fact that the keys generated by the store are content based. This means that if two fragments have the same key this means that their contents are exactly the same. Fragments with different content produce different keys.

An alternative implementation of an update algorithm could be to compare the original fragments with the fragments resulting after updating the XML entity and reifying it. Although this implementation does not depend on the store feedback, is more expensive than the previous, since fragment comparison is performed.

## 8.4 Measurements

This section describes various tests performed using the XBase framework as a storage infrastructure. For time experiments the numbers resulted are seconds, whereas for space experiments the numbers are bytes. It should be noted that each number is the average of performing the experiment 10 times.

The first experiment was to measure time and space for storing and retrieving a simple XML document (XBaseMembers), that contains personal details of 5 people. The resulting numbers are shown in Table 2, The first column contains the results when user specified granularity is applied, whereas the second column contains the results when the default granularity is applied (by default that granularity is the whole XML document). Finally the third column contains the results of performing store and retrieve without using the XBase framework, that is simple file storage and retrieval functionality provided by the Java API.

| | | With granularity | Without granularity | File storage |
|---|---|---|---|---|
| | **XBaseMembers (Time)** | | | |
| 1 | reify | 1.141 | 1.061 | |
| 2 | everything except reify (resolve name, put) | 0.772 | 0.381 | |
| 3 | Total storage | 1.913 | 1.442 | 0.130 |
| 4 | reflect | 0.451 | 0.100 | |
| 5 | everything except reflect (resolve name, put) | 0.050 | 0.040 | |
| 6 | Total retrieval | 0.501 | 0.140 | 0.050 |
| 7 | Reflect & reify | 1.592 | 1.161 | |
| 8 | everything except reflect & reify (resolve name, put) | 0.822 | 0.421 | |
| 9 | Total storage and retrieval | 2.414 | 1.582 | 0.180 |
| | **XBaseMembers (Space)** | | | |
| 10 | Local store | 3281 | 880 | |
| 11 | Root store rep | 864 | 447 | |
| 12 | Namer rep | 1960 | 464 | |
| 13 | Total storage required (bytes) | 6105 | 1791 | 1468 |
| 14 | No of fragments created | 10 | 2 | 2 |

**Table 2:** Measuring time and space for storing and retrieving a simple XML document

Lines 1-3 contain the results for storing the XML document, whereas lines 4-6 contain the results for retrieving the XML document. The totals for performing storage and retrieval are included in lines 7-9.

Lines 10-13 show the amount of space required in each case. Note that in the XBase framework, a file is required to store a Local Store representation (line 10), another file is required to store the Root Store representation (line 11) and finally another file is required to store the Namer representation (line 12). Finally, line 14 includes the numbers of XML fragments created in each case.

The same measuring criteria were used to perform a similar experiment using a much bigger XML document (Othello), which contains the verses of the play. The results we got for that XML document are shown in Table 3:

| | | With granularity | Without granularity | File storage |
|---|---|---|---|---|
| | **Othelo (Time)** | | | |
| 1 | reify | 6.489 | 4.396 | |
| 2 | everything except reify (resolve name, put) | 97.230 | 2.574 | |
| 3 | Total storage | 103.719 | 6.970 | 0.821 |
| 4 | reflect | 3.210 | 2.560 | |
| 5 | everything except reflect (resolve name, get) | 116.220 | 1.917 | |
| 6 | Total retrieval | 119.430 | 4.477 | 0.792 |
| 7 | Reflect & reify | 9.699 | 6.956 | |
| 8 | everything except reflect & reify (resolve name, put) | 213.450 | 4.491 | |
| 9 | Total storage and retrieval | 223.149 | 11.447 | 1.613 |
| | **Othelo (Space)** | | | |
| 10 | Local store | 443000 | 243287 | |
| 11 | Root store rep | 62700 | 447 | |
| 12 | Namer rep | 144000 | 432 | |
| 13 | Total storage required (bytes) | 649700 | 244166 | 242216 |
| 14 | No of fragments created | 1183 | 2 | 2 |

**Table 3:** Same measuring criteria for a bigger XML document

The experiments described above store and retrieve XML documents without taking into consideration the fact that some of the fragments might be updated. A separate experiment was performed, using the XBaseMembers XML document, which measures the time for two different techniques of update and for three different kinds of updates. The results are shown in Table 4.

There are three kinds of update defined in this experiment. The first is to edit an existing fragment (column 1), the second is to add a new fragment (column 2) and the third is to delete an existing fragment (column 3).

The two different update techniques are included in the rows. This is to decide whether an XML fragment is an updated version of an existing fragment in the store or not.

The first technique(lines 1-3) presents the fragments to the store and if the fragment exists it does not add it. In the second technique (lines 4-7) the decision is made before the fragments are presented to the store. This is done by comparing the original fragments with the fragments resulted from performing any kind of update shown in the columns. This measurement includes the time it takes to compare those fragments.

| | Edit | Addition | Deletion |
|---|---|---|---|

| | XBaseMembers | | | |
|---|---|---|---|---|
| 1 | Update by presenting to store | 0.130 | 0.201 | 0.140 |
| 2 | reify | 0.060 | 0.060 | 0.070 |
| 3 | everything except reify | 0.070 | 0.141 | 0.070 |
| | | | | |
| 4 | Update by comparing source | 0.110 | 0.181 | 0.110 |
| 5 | reify | 0.050 | 0.060 | 0.050 |
| 6 | compare | 0.010 | 0.010 | 0.010 |
| 7 | everything except reify & compare | 0.050 | 0.111 | 0.050 |

**Table 4:** Experiment measuring update

# 9 Persistence of Objects

This section addresses various issues related to the persistence of objects documents using the XBase framework (A description of which is provided elsewhere). It mainly refers to the way that object graphs are flattened. The technique used for casting XML document can be used, with some modifications, to cast object/object graphs. In this document we only describe the extra dimensions we should take into account to successfully perform the casting process.

## 9.1 Object Casting

The easiest way to provide object casting tools is to make use of existing technology. One possible way of achieving that is to make use of standard Java serialisation. This technique requires classes to implement interface Serializable.
Another alternative is to serialise the object graph using the standard XML API included in JDK 1.4.0 and then cast the resulting XML document using the XML caster described elsewhere. However, this technique has one main drawback: the XML serialiser can only serialise Java Beans.
Converting objects into java beans or into Serializable objects is a complicated process that require class rewriting and copying of objects, which means that we may loose information about identity.
An alternative to both is to borrow the technique we used for XML casting and modify it accordingly to cast objects. This technique is what we describe in the rest of this document.

### 9.1.1 Policies

The set of policies defined elsewhere (XML persistence document) apply for the case of casting objects. In particular, granularity could be specified per object, per object graph or could be user customised. The simplest case is to specify granularity per object.
The representation format could be in XML format. As for the reference format a name could be used to reference the representation of another object.
The extra dimensions when casting objects are implementation policies:

- **Object State Storing:** this policy has two branches. The first specifies the technique used for object introspection in order to extract the state of an object. Possible alternatives are:
    - Reflection: the limitations here are related to the native methods. It does seem to have any limitations as far as private members is concerned.
    - User specified technique.

  The second sub-policy specifies the information stored within the representation of an object. Possible alternatives are:
    - Its state only: this can be used in the case of having a system that deals with class loading separately.
    - Its state and any relevant code (e.g. in the case of Java this could include the fields and the bytes of the class). The representation of the code can be kept either in the same place as the representation of the

object or kept separately possibly in a different location, in which case the representation of the object should contain a reference to the representation of the corresponding code.

- **Support for Object Updating:** as in the case of casting XML documents, this is to specify the way that the system discovers the differences between different versions of objects. Possible alternatives are:
  - Modify the VM in order to make objects dirty when they are updated. This is not suitable if interoperability is required.
  - Don't modify the VM but keep externally information about "dirty" objects. In order to achieve that techniques such as source code transformation, or byte code transformation may be used. This is not suitable if simplicity is required, that is hide as much as possible from the user.
  - The caster is in charge to decide which objects have been updated. This is done in a similar way to XML documents (this is described elsewhere).

### 9.1.2 A Proposed Object Casting Technique

In this section we describe a particular implementation of a caster. Interface Caster, defined elsewhere, should be modified accordingly since reifying an object may result in more than one representation. Similarly, the reflect operation should be able to accept the result of the reification process, that is a set of representation fragments. Therefore, the original definition of interface caster described elsewhere should be modified to the following (t is the object):

**reify**: *t* -> set[*BitString*] *or error*
**reflect**: set[*BitString*]-> *t or error*

#### 9.1.2.1 Description of the Algorithms

Below we describe the algorithms for both operations. The policies that guide the particular implementation are:

- the representation is in XML format,
- the reference is a name and the namer used to resolve it is in the form of a root namer in a fixed location.

The algorithm for the reify operation is shown in Figure 43:

- If object to be reified ==null then throw an error
- For each object in the object graph {
  - If we have visited that object then do nothing else record that we have visited the object to avoid a cycle.
  - Create an XML representation skeleton (XML-SKEL)
  - Use reflection to extract the fields and their values
  - For each field {
    - generate an XML representation (FIELD-SKEL)
    - if there is a reference to another object then create a reference that has the general form <XbaseRef ref="(SYS-ID)" />.
  - }
  - if we have the class of that particular object then {

- Create a code XML skeleton (CODE-SKEL)
        - Create a reference that has the general form as shown above.
    - }
- }

**Figure 65:** An algorithm for the *reify* operation

The skeleton for an object's representation (XML-SKEL) has the following general form:

```
<XbaseName ref="(SYS-ID)">
        <fields>
                ……
        </fields>
        <code> (CODE-ID) </code>
</XbaseName>
```

The skeleton XML generated for each of the fields (FIELD-SKEL) is shown bellow:

```
<field1>
        <name> … </name>
        <type> … </type>
        <value> (either the value itself if it is primitive or a reference </value
</field1>
```

The sheleton XML generated for code (CODE-SKEL) is shown bellow:

```
<XbaseName ref="(CODE-ID)">
        <code>
                <className> (The name of the class) </className>
                <bytes> (The bytes) </bytes>
        </code>
</XbaseName>
```

SYS-ID and CODE-ID are generated by the system. They have the general form:

- SYS-ID: (Node IP address)+(Object's class full package name)+(counter)
- CODE-ID: (Node IP address)+"class_"+(Class full name) + (counter)

For example:

- SYS-ID: 192.0.0.1-person-1
- SYS-ID: 192.0.0.1-class_person-1

An algorithm for the reflect operation is shown in Figure 44:

- For each given fragment {
    - If it is not well formed then extract the name and the document that is encapsulated, and record that tuple in *tuples*.
    - If it represents a class then extract the bytes and load the class.
- }
- if there is at least one invalid fragment then throw an error
- if we have visited that representation then do nothing else record it to avoid cycles.
- Create a java object (object).

- For each reference in the fragment (XBaseRef element) {
    - Extract the name from the reference
    - If there is a fragment among the given that is identified by a name that is equal with the name included in the reference then keep it for later use (variable *fragment*)
    - else if the root namer contains a binding with a name that is equal to the name included in the reference element then {
        - Find the corresponding key by resolving the name in the root namer
        - Present that key to the root store.
        - If there is a fragment as a result then keep it for later use (variable *fragment*)
        - else throw an error.
        - Replace the reference tag with the element encapsulated in *fragment*
    - } else throw an error
    - Create a java object from that fragment.
    - Update the field value included in object.
    - Recursively scan the children of the element for references and perform the same task.
- }
- Return the object

**Figure 66:** An algorithm for the *reflect* operation

#### 9.1.2.2 An Example

An example of implementing that technique is shown below. We assume that we have two instances of class *Person* and *Address* respectively as live objects but not persistent on host **ouzo**. Object *Person* has a public field *name* with value *"Vangelis"* and another field *address* that references an instance of class *Address* that has a public field *address* with value "*St Andrews*". We want to make them persistent by producing an XML representation for each one. The representation for the instance of class **Person** will look like this:

```
<XbaseName ref="person1">
	<fields>
		<field1>
			<name>name</Name>
			<type>java.lang.String</Type>
			<value>Vangelis</Value>
		</field1>
		<field2>
			<name>address</Name>
			<type>Address</Type>
			<value>
				<XbaseRef ref="192.0.0.1-address1" />
			</value>
		</field2>
	</fields>
	<code> 192.0.0.1-class_Person-1 </code>
</ XbaseName >
```

The representation for the instance of class *Address* will look like this:

```
<XbaseName ref=”192.0.0.1-address1”>
	<fields>
		<field1>
			<name>name</Name>
			<type>java.lang.String</Type>
			<value>St Andrews</Value>
		</field1>
	</fields>
	<code> 192.0.0.1-class_Address-1 </code>
</ XbaseName >
```

The representation for class *Person* is illustrated below (The representation of class *Address* will have similar format):

```
<XbaseName ref=”192.0.0.1-class_Person-1”>
	<code>
		<className> Person </className>
		<bytes> [bytes read from a file] </bytes>
	</code>
</ XbaseName >
```

Having those representations we may put them in a store (possibly the root store) and we get back keys. The resulting keys are bound to names plus timestamps and are placed in the root namer. Newly created objects are separated from objects resulting from the process of reflection by making use of an external mapping between the latter and Ids (this is done in the same way as in the case of XML document in order to provide identity).

Assuming that we perform an update to the instance of class *Person*, the steps we have to follow when putting the representations in the store are similar to the steps that have to be followed in the case of updating XML documents:

- Produce a representation for each object in the object graph. This will look similar to the representations shown above.
- Attempt to put the representations in the store. For unchanged objects this will not have any effect. However, for modified objects this will put the new representation in the store and will return a new key.
- Bind that key with the name of the original object and a new timestamp. That binding will be placed in the root namer (this automatically creates a new object).

The caster deals with cyclic object graphs/representations by keeping track of the objects/representations been traversed. This implies that every time a new object/representation is about to be created, the system checks first whether this has already been done or not.